\documentclass[lettersize,journal]{IEEEtran}

\usepackage{amsmath,amsfonts}
\usepackage[ruled, linesnumbered]{algorithm2e}

\newcommand{\mycommentstyle}[1]{\color[HTML]{000000}{\small #1}}
\SetKwComment{Comment}{\mycommentstyle{// }}{}

\usepackage{array}
\usepackage{textcomp}
\usepackage{stfloats}
\usepackage{url}
\usepackage{verbatim}
\usepackage{graphicx}
\usepackage{cite}

\usepackage{tikz}
\usepackage{filecontents}
\usepackage{subcaption}
\usepackage{caption}
\usepackage{enumitem}
\usepackage{multirow}
\usepackage{booktabs}
\usepackage{tabularx}
\usepackage{lipsum}
\usepackage{float}  
\usepackage{xcolor}

\begin{document}

\title{Semantic-Aware Caching for Efficient Image Generation in Edge Computing}
\author{Hanshuai Cui, Zhiqing Tang,~\IEEEmembership{Member,~IEEE,} Zhi Yao, Weijia Jia,~\IEEEmembership{Fellow,~IEEE}, and Wei Zhao,~\IEEEmembership{Fellow,~IEEE}
\IEEEcompsocitemizethanks{

\IEEEcompsocthanksitem Hanshuai Cui is with School of Artificial Intelligence, Beijing Normal University, Beijing 100875, China, and also with Institute of Artificial Intelligence and Future Networks, Beijing Normal University, Zhuhai 519087, China. E-mail: hanshuaicui@mail.bnu.edu.cn
\IEEEcompsocthanksitem Zhiqing Tang is with Institute of Artificial Intelligence and Future Networks, Beijing Normal University, Zhuhai 519087, China. E-mail: zhiqingtang@bnu.edu.cn
\IEEEcompsocthanksitem Zhi Yao is with School of Artificial Intelligence, Beijing Normal University, Beijing 100875, China, and also with the Institute of Artificial Intelligence and Future Networks, Beijing Normal University, Zhuhai 519087, China. E-mail: yaozhi@mail.bnu.edu.cn.
\IEEEcompsocthanksitem Weijia Jia is with the Institute of Artificial Intelligence and Future Networks, Beijing Normal University, Zhuhai 519087, China and also with Guangdong Key Lab of AI and Multi-Modal Data Processing, Beijing Normal-Hong Kong Baptist University, Zhuhai 519087, China. E-mail: jiawj@bnu.edu.cn
\IEEEcompsocthanksitem Wei Zhao is with Shenzhen University of Advanced Technology, Shenzhen 518055, China. E-mail: zhao.wei@siat.ac.cn
\IEEEcompsocthanksitem (\textit{Corresponding authors: Zhiqing Tang and Weijia Jia.})
}
}



\maketitle

\begin{abstract}
  Text-to-image generation employing diffusion models has attained significant popularity due to its capability to produce high-quality images that adhere to textual prompts. However, the integration of diffusion models faces critical challenges into resource-constrained mobile and edge environments because it requires multiple denoising steps from the original random noise. A practical way to speed up denoising is to initialize the process with a noised reference image that is similar to the target, since both images share similar layouts, structures, and details, allowing for fewer denoising steps. Based on this idea, we present CacheGenius, a hybrid image generation system in edge computing that accelerates generation by combining text-to-image and image-to-image workflows. It generates images from user text prompts using cached reference images. CacheGenius introduces a semantic-aware classified storage scheme and a request-scheduling algorithm that ensures semantic alignment between references and targets. To ensure sustained performance, it employs a cache maintenance policy that proactively evicts obsolete entries via correlation analysis. Evaluated in a distributed edge computing system, CacheGenius reduces generation latency by 41\% and computational costs by 48\% relative to baselines, while maintaining competitive evaluation metrics. 
\end{abstract}

\begin{IEEEkeywords}
Diffusion models, edge computing, image generation, semantic caching, latency optimization.
\end{IEEEkeywords}

\section{Introduction}
\IEEEPARstart{T}{he} rapid advancement of Artificial Intelligence-Generated Content (AIGC) has revolutionized on-demand visual content creation for mobile and edge platforms, enabling high-quality image generation from textual prompts \cite{singh2016creating}. In practice, teachers can instantly generate lesson-specific illustrations on tablets during mobile teaching scenarios \cite{ali2024picture}, intelligent vehicles can execute real-time visual path planning via in-vehicle edge systems \cite{du2023enabling}, and users can create AR-adaptive immersive objects on wearable devices in the metaverse \cite{du2024diffusion}. These capabilities span domains from media production to interactive entertainment, enhancing creativity and efficiency across various industries. A report by Gartner predicts that by 2026, 90\% of all digital content will be AI-generated \cite{Gartner}.

\begin{figure}[t]
  \centering
  \includegraphics[width=0.9\linewidth]{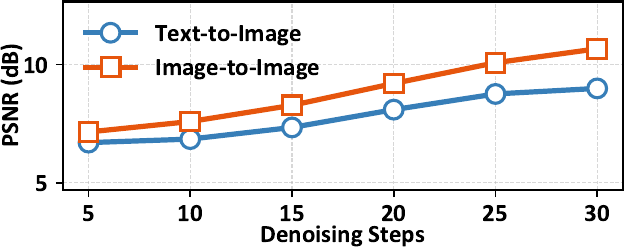}
  \caption{The evolution of PSNR during text-to-image and image-to-image denoising processes}
  \label{fig:psnr}
\end{figure}

Early text-to-image models, such as Variational Autoencoders (VAEs) \cite{kingma2013auto} and Generative Adversarial Networks (GANs) \cite{goodfellow2014generative}, suffered from blurry outputs, training instability, and limited text-image alignment. Diffusion Models (DMs) \cite{ho2020denoising,dhariwal2021diffusion} address these limitations by enabling stable training, high-quality generation via stepwise denoising, and improved capture of complex data distributions, resulting in richer diversity and finer details. Although DMs excel in performance, their image quality heavily relies on denoising steps. More steps improve quality but increase GPU consumption and latency, whereas fewer steps degrade quality, yielding a critical efficacy-usability trade-off \cite{zhang2023redi, ma2024deepcache,agarwal2024approximate}. Deploying DMs at the edge offers unique benefits, such as preserving sensitive data, avoiding cloud routing delays, and processing closer to end-users \cite{xu2024unleashing,khalili2024lightpure}. However, the limited computational capacity of edge devices severely limits denoising iterations, further compromising the output quality.

\begin{figure}[t]
  \centering
  \begin{subfigure}[b]{0.5\textwidth}
    \centering
    \includegraphics[width=\textwidth]{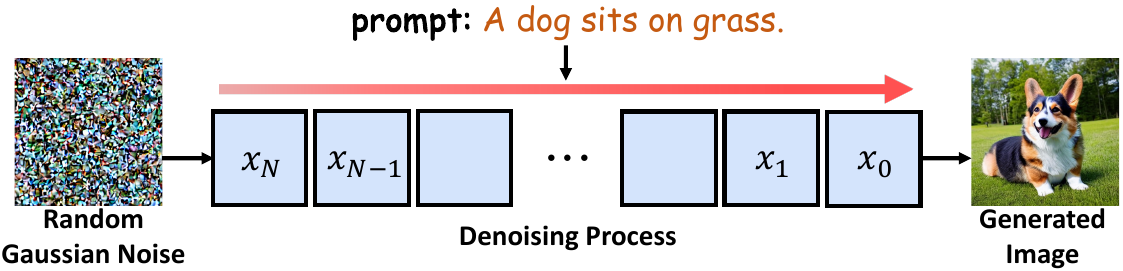}
    \caption{}
    \label{fig:dog}
  \end{subfigure}
  \hfill 
  \begin{subfigure}[b]{0.5\textwidth}
    \centering
    \includegraphics[width=\textwidth]{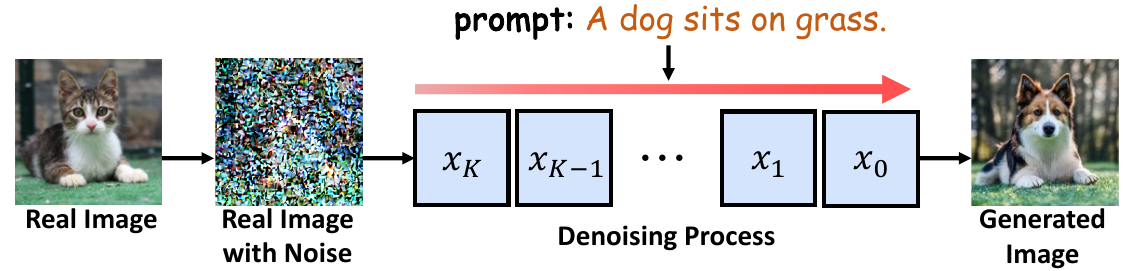}
    \caption{}
    \label{fig:cat}
  \end{subfigure}
  \caption{Comparison of text-to-image and image-to-image workflows. (a) Text-to-image. (b) Image-to-image.}
  \label{fig:img2img}
\end{figure}

Given the resource constraints of edge devices, we focus on a promising topic: \textit{How to reduce image generation latency while ensuring output quality?} Prior work accelerates DM sampling by reducing parameters through distillation, pruning, or quantization \cite{wang2024towards,chu2024qncd,yang2024pruning,zhang2024laptop,kang2024distilling,zhou2024simple}, but these approaches require retraining. Post-training acceleration methods can be implemented without altering the original model \cite{ma2024deepcache,zhang2023redi,wimbauer2024cache,agarwal2024approximate}; however, they typically either incur cold-start latency or depend on architectural changes. Our study demonstrates that image-to-image generation requires significantly fewer denoising steps than text-to-image generation. As shown in Figure \ref{fig:psnr}, the Peak Signal-to-Noise Ratio (PSNR), a metric correlated with the discrepancy between original and noise images, evolves more rapidly in image-to-image denoising. Higher PSNR values correspond to reduced noise levels \cite{thomos2005optimized}. Notably, the image-to-image process achieves a PSNR at 20 iterations that surpasses the text-to-image at 30 iterations. 
When reference images (e.g., images sharing analogous layouts, shapes, or annotations) align closely with the target output, image-to-image achieves high-quality results with fewer steps, thereby reducing computation. Figure \ref{fig:img2img} contrasts text-to-image and image-to-image. In Figure \ref{fig:dog}, random Gaussian noise undergoes $N$ denoising steps to generate an image matching the text prompt. In Figure \ref{fig:cat}, a noise-corrupted reference image undergoes fewer steps ($K < N$), guided by both the reference latent structure and the text prompt. 

Although reference-guided image generation can speed up generation, deploying this module on edge platforms introduces two critical challenges. First, the task is inherently personalized: diverse user groups expect uniquely tailored outputs, while edge platforms must operate with limited, heterogeneously distributed resources. \textit{The first challenge lies in how to quickly find reference images that closely match each specific generation request under resource-constrained environments.} Second, although caching reference images improves efficiency, the limited storage capacity of edge nodes conflicts with the need to serve a wide variety of requests. \textit{The second challenge is how to dynamically adjust cached images based on evolving user demands to prevent limited resources from being wasted on invalid or irrelevant data.}

\begin{figure}[t]
  \centering
  \includegraphics[width=\linewidth]{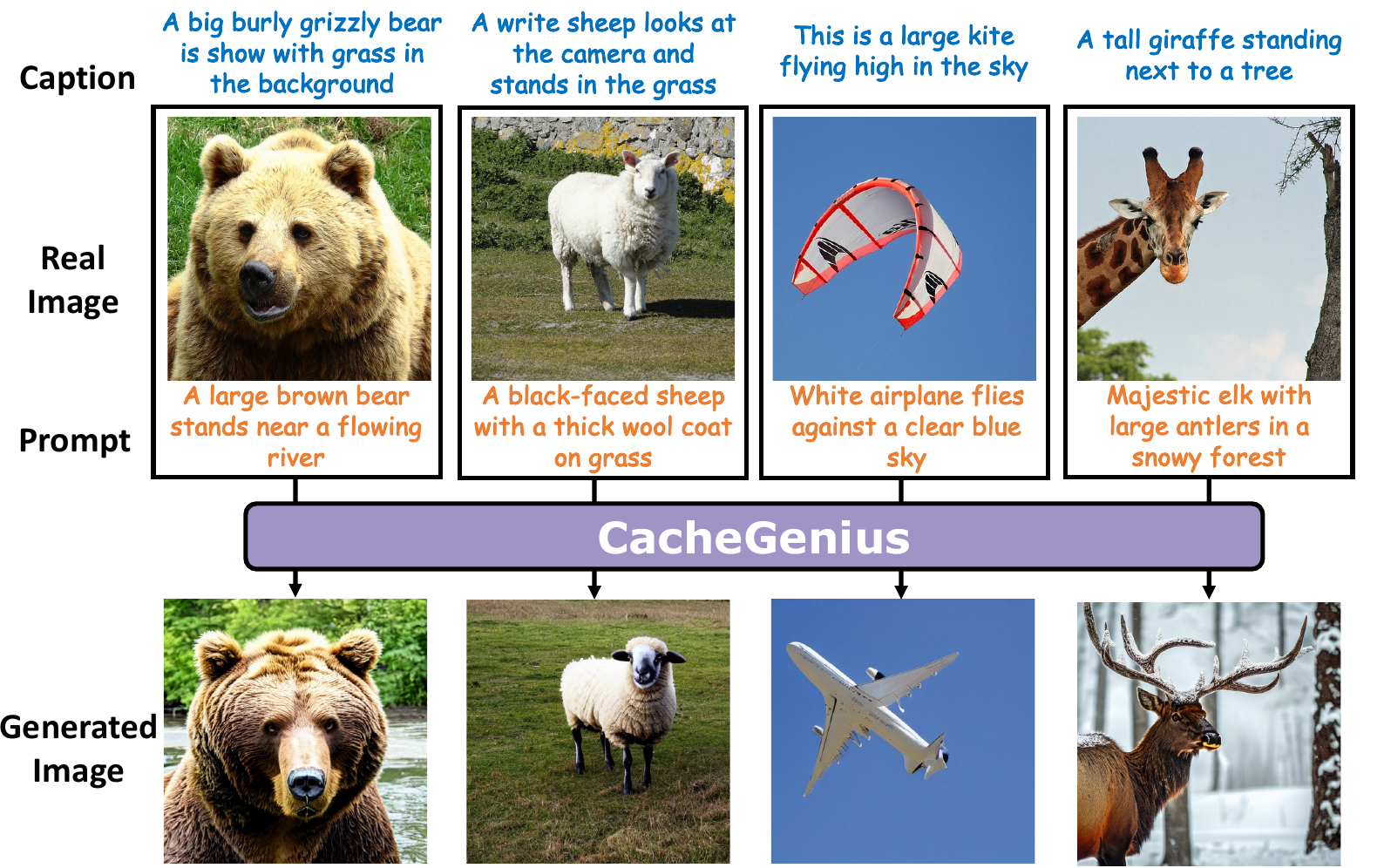}
  \caption{Images generated using CacheGenius}
  \label{fig:CacheGenius}
\end{figure}

In this paper, we introduce CacheGenius, a hybrid image generation system for latency-sensitive applications. CacheGenius accepts user text prompts and generates images by combining text-to-image and image-to-image workflows grounded in cached reference images. By leveraging these references, the system preserves output quality while reducing computation. It further strategically schedules tasks across a distributed edge environment and employs semantic-aware retrieval from vector databases (VDBs) to minimize both latency and cost. Figure \ref{fig:CacheGenius} demonstrates examples using real captioned images from several datasets, illustrating the reference image-based generation workflow. First, CacheGenius introduces a reference image-driven generation system that sharply reduces denoising latency. Second, a semantically classified storage scheme and a request-scheduling algorithm are proposed to retrieve relevant reference images efficiently. Third, a novel Least Correlation Used (\texttt{LCU}) cache maintenance policy is designed to dynamically update cached images based on their semantic alignment with current query distributions. Deployed in heterogeneous edge computing environments, CacheGenius reduces generation latency by 41\% and computational costs by 48\% compared to baselines, while outperforming them across various image evaluation metrics. 

The main contributions are summarized as follows.
\begin{itemize}[noitemsep, leftmargin=*, topsep=0pt]
  \item We propose a hybrid image generation system that integrates text-to-image and image-to-image workflows via cached reference images. This system significantly reduces generation latency and cost, particularly effective in resource-limited edge scenarios.
  \item We present a semantic-aware classified storage method and a request-scheduling algorithm that leverages multimodal feature matching to rapidly retrieve reference images aligned with user prompts.
  \item We design the \texttt{LCU}, a novel cache maintenance policy that dynamically updates cached images. This policy ensures high cache hit rates by prioritizing frequently accessed reference images while removing obsolete data.
  \item We deploy and evaluate our system in a distributed edge environment. Our implementation achieves an approximately 41\% reduction in image generation latency compared to baselines like Stable Diffusion.
\end{itemize}

\section{Related Works}
\textbf{Text-to-image models}: Most industrial text-to-image systems, such as DALL-E \cite{dalle}, Midjourney \cite{midjourney}, and Adobe Firefly \cite{firefly}, remain proprietary, limiting accessibility for customization. DMs \cite{ho2020denoising,dhariwal2021diffusion} have become the cornerstone of open-source text-to-image generation, with accelerated sampling methods like DDIM \cite{Song2022} and latent-space optimizations \cite{Rombach2022} reducing denoising steps. Existing acceleration techniques for DMs rely heavily on architectural changes: distillation simplifies networks \cite{kang2024distilling,zhou2024simple,salimans2024multistep}, pruning removes redundant parameters \cite{yang2024pruning,zhang2024laptop,castells2024ld}, quantization reduces precision, and others \cite{wang2024towards,chu2024qncd,huang2024tfmq}. While these methods reduce denoising costs (e.g., SD-Tiny \cite{kim2023bksdm} cuts latency by 40\%), they degrade output quality. CacheGenius diverges by exploiting semantic similarity between prompts and cached references to reduce denoising steps without altering the model architecture, preserving output quality while achieving generation speedups.

\textbf{Caching for generative models}: There has been substantial research focused on applying caching mechanisms in DMs. These studies typically optimize the denoising process by storing and reusing intermediate computational results \cite{liu2024efficient,wimbauer2024cache,zhang2023redi}. However, most of these methods require pre-running the system for a period of time to initialize the cache, whereas CacheGenius eliminates the cold-start latency by utilizing public datasets as cached data. NIRVANA \cite{agarwal2024approximate} is a system leveraging approximate caching to optimize DMs by reusing intermediate denoising states from prior prompts. In addition to cold start issues, NIRVANA also faces the problem that massive intermediate results consume significant storage space. DeepCache \cite{ma2024deepcache} reduces redundant U-Net computations by caching high-level features across timesteps. This method is tightly coupled to specific architectures; however, our image-level approach remains model-agnostic. This enables seamless deployment across different diffusion model variants without modification. Retrieval-based methods \cite{bang2023gptcache,pinecone2023stable} match prompts to cached images via embeddings but ignore structural similarity.

\textbf{AIGC task scheduling}: While AIGC adoption has surged \cite{cao2025survey,Chi2024}, existing systems face challenges in balancing quality, latency, and resource constraints \cite{feng2025resource,zhuansun2025generative,huang2024digital,zhang2023redi,Huang2025}. Efficient scheduling of AIGC workloads is critical for latency-sensitive applications. Du \textit{et al.} \cite{du2024diffusion} introduced an AGOD algorithm to optimize the selection of AIGC service providers in edge networks, addressing resource constraints and personalized user demands in the Metaverse. However, these methods merely focus on decision node selection \cite{hazarika2025generative}, model selection \cite{liang2024resource}, and generation steps \cite{ye2024optimizing}, without optimizing the generation process of DMs.

\begin{figure*}[t]
  \centering
  \includegraphics[width=0.85\textwidth]{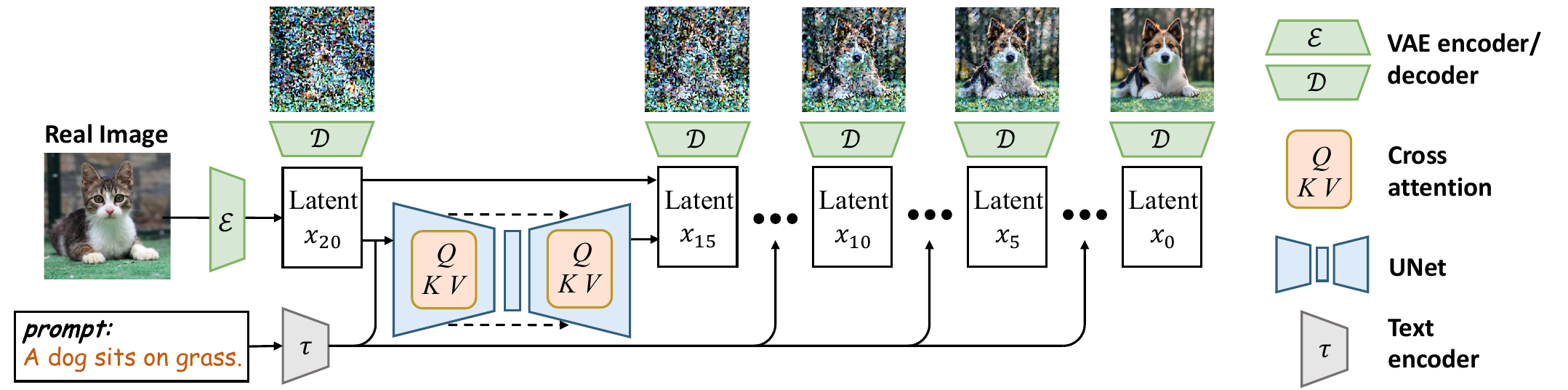}   
  \caption{Workflow of image-to-image with diffusion model}
  \label{fig:DDIM}
\end{figure*}

\section{Preliminary}
\subsection{Diffusion Models}

DMs represent a significant advancement in AIGC, framing image generation as an iterative denoising process. A typical DM framework, the Denoising Diffusion Probabilistic Model (DDPM) \cite{ho2020denoising,dhariwal2021diffusion}, defines a Markov chain to progressively add Gaussian noise to images and then learn to reverse this process, generating samples from the noise.

\textbf{Forward and reverse processes}: Consider the data $\mathbf{x}_0 \sim q(\mathbf{x})$ sampled from a real distribution. The forward diffusion process incrementally adds Gaussian noise over $T$ steps, producing a sequence $\mathbf{x}_1, \dots, \mathbf{x}_T$. As $t$ increases, $\mathbf{x}_t$ progressively loses distinguishable features, asymptotically approaching an isotropic Gaussian distribution when $T \to \infty$. With the Markov chain assumption, it is expressed as:
\begin{equation}
\begin{aligned}
  q(\mathbf{x}_{1:T} \vert \mathbf{x}_0) &= \prod^T_{t=1} q(\mathbf{x}_t \vert \mathbf{x}_{t-1}),\\
  q(\mathbf{x}_t \vert \mathbf{x}_{t-1}) &= \mathcal{N}(\mathbf{x}_t; \sqrt{1 - \beta_t} \mathbf{x}_{t-1}, \beta_t\mathbf{I}),
\end{aligned}
\end{equation}
where $q(\mathbf{x}_t \vert \mathbf{x}_{t-1})$ is the posterior probability. The step sizes are controlled by a noise schedule $\mathbf{\beta}_1, \dots, \mathbf{\beta}_T$. The reverse process samples from $q(\mathbf{x}_{t-1} \vert \mathbf{x}_t)$ to reconstruct data from noise $\mathbf{x}_T \sim \mathcal{N}(\mathbf{0}, \mathbf{I})$. At each step, the model predicts noise components through:
\begin{equation}
\mathbf{x}_{t-1} = \frac{1}{\sqrt{\alpha_t}} \left( \mathbf{x}_t - \frac{\beta_t}{\sqrt{1-\bar{\alpha}_t}} \epsilon_\theta(\mathbf{x}_t, t) \right) + \sigma_t z,
\end{equation}
where $\alpha_t = 1-\beta_t$, $\bar{\alpha}_t = \prod_{s=1}^t \alpha_s$, and $z \sim \mathcal{N}(0, \mathbf{I})$. The function approximator $\epsilon_\theta$ estimates the noise $\boldsymbol{\epsilon}$ from $\mathbf{x}_t$.

\subsection{Accelerating Diffusion Models}

The generation process in DDPM is computationally intensive due to its reliance on full Markov chain iterations, typically requiring hundreds or thousands of steps.

\textbf{Accelerated sampling}: Denoising Diffusion Implicit Model (DDIM) \cite{Song2022} does not have to follow the entire chain $t=1,\dots,T$, but rather a subset of steps. The DDIM update step is defined as follows:
\begin{equation}
  \begin{split}
    \mathbf{x}_{t-1} &= \sqrt{\bar{\alpha}_{t-1}} \left( \frac{\mathbf{x}_t - \sqrt{1 - \bar{\alpha}_t} \epsilon^{(t)}_\theta(\mathbf{x}_t)}{\sqrt{\bar{\alpha}_t}} \right) \\
    &\quad + \sqrt{1 - \bar{\alpha}_{t-1} - \sigma_t^2} \epsilon^{(t)}_\theta(\mathbf{x}_t) + \sigma_t\boldsymbol{\epsilon}_t,
  \end{split}
\end{equation}
where $\boldsymbol{\epsilon}_t \sim \mathcal{N}(0, \mathbf{I})$ is standard Gaussian noise independent of $\mathbf{x}_t$. This formulation allows training with arbitrary forward steps while sampling from a reduced subset. We adopt this accelerated approach in our work.

\textbf{Latent variable space}: Latent Diffusion Model (LDM) \cite{Rombach2022} reduces computational costs by operating in a compressed latent space learned by an autoencoder. The autoencoder encodes images into lower-dimensional latent $\mathbf{z} = \mathcal{E}(\mathbf{x})$, and the diffusion process occurs in this space. Stable Diffusion \cite{Rombach2022} is a specific implementation and extension of latent diffusion, developed and open-sourced by Stability AI \cite{Stability}, with a primary focus on text-to-image generation.

\subsection{Text-to-Image and Image-to-Image}
While both generations utilize DMs, the noise initialization strategies of text-to-image and image-to-image are fundamentally different:

\textbf{Text-to-image}: Initializes from pure Gaussian noise $\mathbf{x}_T \sim \mathcal{N}(0, \mathbf{I})$, enabling unconstrained generation while necessitating precise semantic alignment between textual prompts and visual outputs \cite{Rombach2022}. This stochasticity can produce diverse but occasionally unstable results, particularly in complex scenes.

\textbf{Image-to-image}: Stochastic Differential Editing (SDEdit) \cite{Meng2022} is an image-to-image generation technique based on DM. Unlike traditional DMs, which start generation from random Gaussian noise, SDEdit treats input images as intermediate states in the diffusion process, enabling generation from ``partial noise” to target images by controlling the noise injection strength. Given the input image, SDEdit first injects noise through a partial forward diffusion process:
\begin{equation}
  \mathbf{x}_t = \sqrt{\bar{\alpha}_t} \mathbf{x}_0 + \sqrt{1 - \bar{\alpha}_t} \boldsymbol{\epsilon}, \quad \boldsymbol{\epsilon} \sim \mathcal{N}(\mathbf{0}, \mathbf{I}),
  \end{equation}
where $t$ controls noise strength. Subsequent denoising produces outputs that balance flexibility with similarity to $\mathbf{x}_0$. 
Figure \ref{fig:DDIM} illustrates the image-to-image generation workflow. The process starts with a given prompt and an initial input image. The text encoder first processes the prompt into text embeddings, while the VAE encoder converts the input image into a noisy latent. This noisy latent, denoted as $\mathbf{x}_{20}$, undergoes 20 sequential denoising steps within the UNet architecture. At each step, the UNet leverages cross-attention mechanisms to condition the latent on the token embeddings. The UNet predicts the noise present in the current latent, subtracts it iteratively, and generates progressively refined latent representations. After completing the denoising steps, the final latent $\mathbf{x}_{0}$ is decoded by the VAE decoder into the output image, generating visual details aligned with both the textual prompt and the structure from the input image.

\section{CacheGenius}
\subsection{Overview}

\begin{figure}[t]
  \centering
  \includegraphics[width=\linewidth]{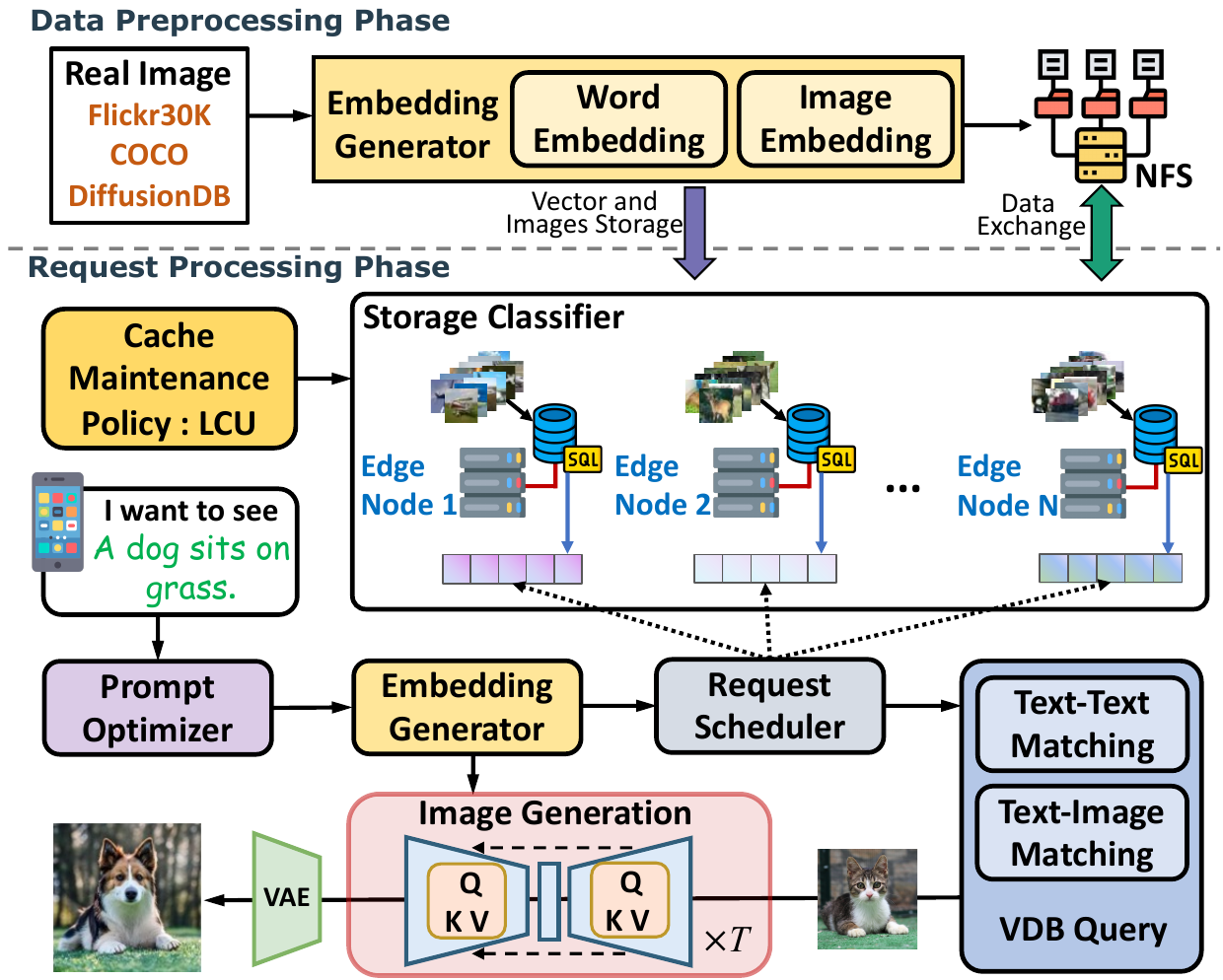}
  \caption{CacheGenius overview}
  \label{fig:architecture}
\end{figure}

Figure \ref{fig:architecture} illustrates the overall architecture of CacheGenius. In the data-preprocessing phase, the system first downloads image datasets and stores them in a distributed Network File System (NFS) to enable shared access across nodes. Each image is paired with a text caption, and both modalities are encoded into 512-dimensional image and text embeddings using the \texttt{embedding-generator} (\S \ref{sec:embedding}). These embeddings are then indexed into VDBs. We deploy a local VDB on each edge node and design a \texttt{storage-classifier} (\S \ref{sec:storage}) that uses K-means clustering to perform unsupervised clustering in the feature space and assign semantically similar vectors to the same node. This similarity-aware placement significantly improves the efficiency of subsequent nearest neighbor retrieval.

In the request-processing phase, CacheGenius employs a hybrid image generation pipeline for efficient image synthesis. When an edge device initiates an image generation task, the \texttt{prompt-optimizer} (\S \ref{sec:opt}) performs semantic analysis of the input prompt $\mathcal{P}$, applies dependency parsing to restructure phrases, and produces structured prompts aligned with the model priors of DMs. The prompt is encoded into a text embedding via the \texttt{embedding-generator}, and the \texttt{request-scheduler} (\S \ref{sec:schedule}) selects the optimal edge node based on the similarity between the prompt embedding and node representation vectors. The \texttt{image-generator} (\S \ref{sec:gen}) queries the VDB on the selected node using multimodal features to retrieve a candidate image set, then computes similarity scores between the prompt embedding and the retrieved images. If the top similarity score exceeds a preset threshold, the corresponding image is used as a reference for image-to-image generation. Otherwise, the text-to-image pipeline is executed; the generated results are stored in NFS, and the VDB is updated. Concurrently, CacheGenius improves output fidelity with a quality-aware priority scheduler: for users requiring higher image quality (e.g., artistic creation or professional research), higher quality generation is prioritized, especially on repeated prompts.

In edge environments, the rapid growth in requests from edge devices poses dual challenges of constrained storage capacity and query efficiency for VDBs. To address this, we propose a cache maintenance policy called Least Correlation Used (\texttt{LCU}) (\S \ref{sec:lcu}). LCU measures the correlation between cached vectors and current query features via a similarity score and maintains a priority queue. It periodically evicts low-correlation vectors at the tail of the queue and synchronously removes the corresponding image files to maintain data consistency. This policy improves storage utilization on edge nodes while sustaining fast, efficient image retrieval.

\subsection{Embedding Generator}
\label{sec:embedding}
The \texttt{embedding-generator} uses the Contrastive Language-Image Pre-Training (CLIP) model \cite{radford2021learning} to achieve cross-modal feature encoding. 
Text and image vectors from both modalities are L2-normalized and mapped into a 512-dimensional latent space, $v_{\text{txt}} \in \mathbf{R}^{512}$ and $v_{\text{img}} \in \mathbf{R}^{512}$, with alignment quantitatively evaluated using cosine similarity. This alignment enables cross-modal similarity measurement, laying the foundation for subsequent semantic retrieval. The alignment between prompts and generated images using different embedding models is evaluated in \S \ref{sec:efficiency}.

\subsection{Storage Classifier}
\label{sec:storage}
The CacheGenius system employs a similarity-guided reference image generation paradigm, with its key innovation lying in an efficient reference image retrieval component. Mainstream computer vision datasets provide a rich candidate resource pool for reference images. 
However, given the massive size of such datasets, distributing them across edge nodes requires dividing the full dataset into subsets. After encoding via the CLIP model, the corresponding vectors are stored in localized VDBs at each edge node to optimize the retrieval efficiency. While traditional storage schemes based on semantic categories (e.g., animals, plants, natural scenes) align with human cognitive habits, image generation tasks rely more heavily on structural similarity (e.g., images sharing analogous layouts, shapes, or annotations) and hold higher reference values. 
For example, generating a bird and an airplane might leverage the same reference image despite belonging to entirely unrelated semantic categories.

\begin{figure}[t]
  \centering
  \begin{subfigure}[b]{0.218\textwidth}
    \centering
    \includegraphics[width=\textwidth]{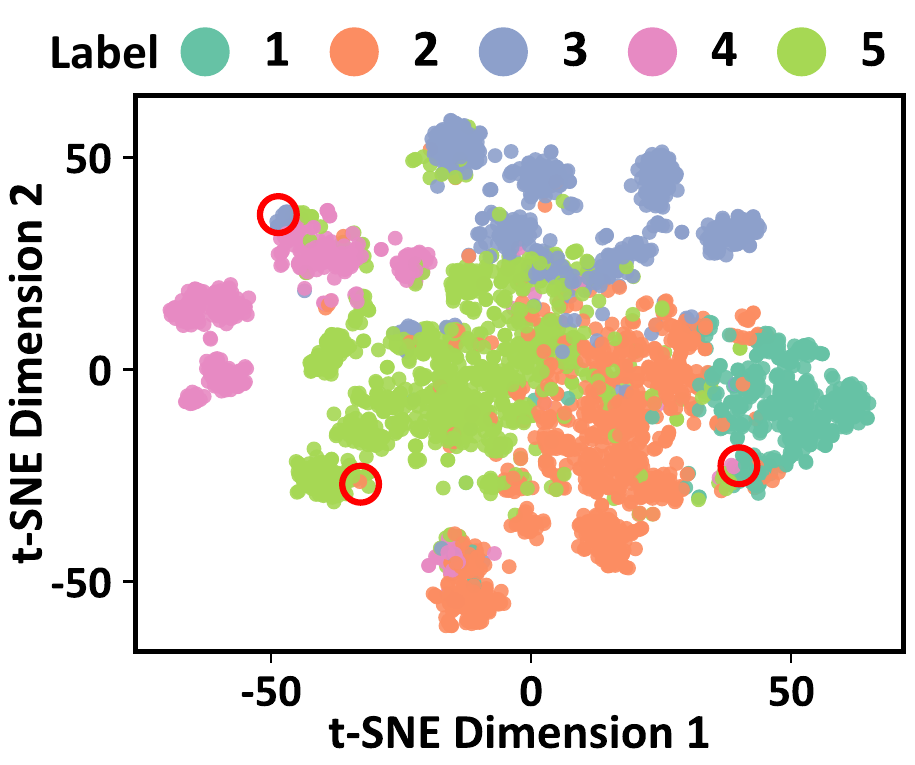}
    \caption{}
    \label{fig:tsne}
  \end{subfigure}
  \hfill 
  \begin{subfigure}[b]{0.25\textwidth}
    \centering
    \includegraphics[width=\textwidth]{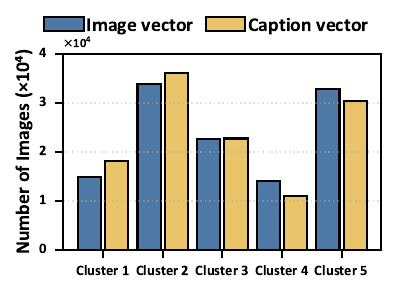}
    \caption{}
    \label{fig:scientific_bar}
  \end{subfigure}
  \caption{(a) T-SNE visualization of data distribution after clustering. (b) Number of images in different clusters after clustering.}
  \label{fig:method}
\end{figure}

To bridge this gap, we propose a clustering-driven \texttt{storage-classifier} module. During the feature extraction phase, the CLIP model encodes images and their corresponding textual descriptions into 512-dimensional vectors. Separate clustering analyses are performed on image and text vectors to reveal potential cross-modal semantic correlations. We adopt K-Means as the foundational clustering algorithm due to its computational efficiency, simplicity, and scalability. The iterative algorithm partitions $n$ data points into $k$ clusters, assigning each point to the nearest cluster centroid to minimize the within-cluster sum of squared errors. The objective function is defined as:
\begin{equation}
J = \sum_{i=1}^k \sum_{\mathbf{x} \in C_i} \|\mathbf{x} - \mathbf{\mu}_i\|^2,
\end{equation}
where $C_i$ is the $i$-th cluster, and $\mathbf{\mu}_i$ denotes the centroid of $C_i$. The squared Euclidean distance is represented by $\|\cdot\|^2$. The number of clusters is set equal to the number of edge nodes in the edge environment. After the clustering algorithm converges, the corresponding vectors in each cluster are stored in the VDB of the associated edge node. Figure \ref{fig:tsne} illustrates the distribution of clustered image vectors. Given the 512-dimensional vector length, we employ t-distributed Stochastic Neighbor Embedding (t-SNE) \cite{van2008visualizing} to project the clustered vectors onto a two-dimensional plane for effective visualization. The analysis reveals that the image vectors form five statistically distinct cluster structures in the latent space. Besides clustering image vectors, we also cluster text vectors. As demonstrated in Figure \ref{fig:scientific_bar}, the results indicate high consistency between text vector clusters and image vector clusters. Considering both cross-modal representation consistency and edge storage efficiency, this study selects image vectors as the classified storage scheme.

In CacheGenius, the NFS serves as the distributed storage component, providing efficient data sharing and centralized management capabilities. NFS enables seamless access to unified storage resources across edge nodes through local area networks (LANs). 
In the operational phase, newly generated images are archived to NFS, serving both as reusable resources for subsequent requests and enabling cross-node content synchronization. This architecture maintains simplicity through a unified storage layer that streamlines cache management operations, which reduces overall storage costs by eliminating data redundancy across edge nodes.

\subsection{Prompt Optimizer}
\label{sec:opt}
In DM, the default weight of each phrase in the prompt is set to 1, with their influence progressively decreasing from left to right based on positional order. These weight distributions significantly impact the visual output generation. Selecting the appropriate order for each phrase enables the generation of high-quality images. Therefore, the \texttt{prompt-optimizer} module is proposed. First, dependency parsing is performed on the prompt using the SpaCy framework \cite{spacy}, splitting it into multiple phrases. The attention weight matrix of the BERT model \cite{devlin2019bert} is employed to determine the relative importance weights of phrases. These phrases are then sorted in descending order of significance and processed into a structured prompt. This module provides more precise emphasis on key elements in target generation than traditional natural language descriptions, enabling fine-grained semantic control.

\subsection{Request Scheduler}
\label{sec:schedule}
\begin{figure}[t]
  \centering
  \includegraphics[width=0.85\linewidth]{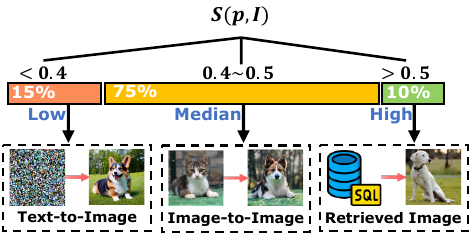}
  \caption{Determine the image generation workflow according to the similarity score}
  \label{fig:value}
\end{figure}

Random request scheduling could fail to retrieve similar reference images, necessitating a full image generation process and incurring additional VDB query latencies. Therefore, the \texttt{request-scheduler} adopts a multimodal feature matching module to allocate user requests to optimal edge nodes.
The scheduler operates by scheduling user requests to edge nodes hosting VDBs that exhibit the highest semantic similarity to their respective prompts. User prompts are encoded using the CLIP model, and feature vectors from edge node VDBs are aggregated to form node-level semantic representation vectors. These representations are computed as the mean average of all vectors contained in the VDBs associated with this node. Then, cosine similarity calculations determine the optimal node matching. The set of edge nodes is defined as $\mathbf{N} = \{n_1,n_2,\dots,n_{|\mathbf{N}|}\}$, where $|\cdot|$ indicates the number of elements in the set. Therefore, given a user prompt vector $v_\mathcal{P} \in \mathbf{R}^{512}$ and node representation vectors $v_{N_i} \in \mathbf{R}^{512}$, the similarity matching score is computed as:
\begin{equation}
S_{match}(v_\mathcal{P},v_{N_i}) = \frac{v_\mathcal{P} \cdot v_{N_i}}{\|v_\mathcal{P}\| \|v_{N_i}\|}.
\end{equation}

Requests are scheduled to the edge node corresponding to $\arg\max_i(S_{match})$, bypassing exhaustive VDB retrieval and substantially reducing query latency. To meet stringent requirements for high-quality images in artistic creation and professional research, the system implements quality-aware priority scheduling. Upon detecting repeated prompt requests, it schedules requests to the highest-performance GPU nodes and executes a text-to-image generation process. This mechanism bypasses VDB queries while ensuring quality and reducing latency. Additionally, historical query caching enables the direct reuse of generated images for highly similar prompts across users, eliminating scheduling and query latencies.

\subsection{Image Generator}
\label{sec:gen}
The \texttt{image-generator} module, as the core of CacheGenius, employs a hybrid image generation system for efficient image generation. 
In edge nodes, the input prompt is initially encoded into vector representations via the \texttt{embedding-generator}. These vectors undergo dual retrieval processes in the VDB through Approximate Nearest Neighbor (ANN) searches, obtaining two top-K candidate sets. Specifically, the ANN search applied to the image vector produces the image-retrieved set $\textbf{I}_{\text{img}}$, while the text vector produces the text-retrieved set $\textbf{I}_{\text{txt}}$. The unified retrieved image set $\mathbf{I}$ is subsequently evaluated for prompt-image relevance using a composite similarity score integrating CLIP Score \cite{radford2021learning} and PicksScore \cite{kirstain2023pick}, defined as:
\begin{equation}
S_{\text{sim}}(\mathcal{P},I_r) = CLIP Score(\mathcal{P},I_r) + PickScore(\mathcal{P},I_r).
\end{equation}

The image with the highest similarity score in the retrieved image set is treated as the reference image $I_r$. Figure \ref{fig:value} shows the image generation workflow for different similarity scores. Scores above 0.5 indicate high relevance, triggering a direct return of retrieved images. Scores below 0.4 suggest irrelevance, which requires text-to-image generation from random noise. The medium relevance range (0.4 to 0.5), which covers most cases, uses reference images for image-to-image generation because they are partially similar to the target outputs.

\begin{algorithm}[h]
  \caption{Similarity Matching and Generation Strategy}
  \label{alg:similarity_matching}
  \KwIn{User prompt $\mathcal{P}$}
  \KwOut{Generated image $I$}
  $v \gets \text{CLIP\_Text\_Encoder}(\mathcal{P})$\;
  $\mathbf{I}_{\text{img}} \gets \text{ANN\_Search}(v, \text{VDB.image\_vectors}, k)$\;
  $\mathbf{I}_{\text{txt}} \gets \text{ANN\_Search}(v, \text{VDB.text\_vectors}, k)$\;
  $\mathbf{I} \gets \mathbf{I}_{\text{img}} \cup \mathbf{I}_{\text{txt}}$ \Comment{Combine retrieved images}
  \For{each image $I_i \in \mathbf{I}$}{
      $S_{sim}(\mathcal{P},I_i) = CLIP Score(\mathcal{P},I_i) + PickScore(\mathcal{P},I_i) $\;
  }
  $I_{\text{r}} \gets \arg\max_{I_i \in \mathbf{I}} S_{\text{sim}}^i$ \Comment{Optimal candidate selection}
  \eIf{$S_{\text{sim}}^{\text{r}} > 0.5$}{
      \Return $I_{\text{r}}$ \Comment{High relevance: direct return}
  }{
      \eIf{$0.4 \leq S_{\text{sim}}^{\text{r}} \leq 0.5$}{
          $I_{\text{gen}} \gets \text{ImageToImage}(\mathcal{P}, I_{\text{r}})$\;
          \Return $I_{\text{gen}}$ \Comment{Medium relevance: generate with reference}
      }{
          $I_{\text{gen}} \gets \text{TextToImage}(\mathcal{P})$\;
          \Return $I_{\text{gen}}$ \Comment{Low relevance: generate from random noise}
      }
  }
\end{algorithm}

Optimizing similarity thresholds in the \texttt{image-generator} module balances output quality and computational efficiency. Lower thresholds speed up VDB reference retrieval, but may reduce image quality with less similar matches. Higher thresholds slow the generation by restricting references, but improve quality through prioritized relevance. The experiment results show that the similarity score for the images generated by the SD-Tiny model \cite{kim2023bksdm} is around 0.5, thereby establishing a similarity threshold of 0.5. Detailed experimental results are presented in \S \ref{sec:efficiency}.

Algorithm \ref{alg:similarity_matching} details the complete similarity matching and generation strategy workflow.

For a request $i$, let $x_i,y_i,z_i \in \{0,1\}$ indicate to directly return the cached image, generate via image-to-image, and generate via text-to-image, respectively. Denote $K$ as the denoising steps for image-to-image workflows and $N$ as the denoising steps for text-to-image workflows ($K < N$). The latency for generating an image is defined as:
\begin{equation}
L_i = t_{\text{retrieve}} + x_i \cdot t_{\text{return}} + y_i \cdot \left( t_{\text{noise}} + K \cdot t_{\text{step}} \right) + z_i \cdot N \cdot t_{\text{step}},
\end{equation}
where $t_{retrieve}$ is the latency to search the VDB for reference images, $t_{return}$ is the latency to transfer the cached image to the user, $t_{noise}$ is the latency to add noise into reference image, and $t_{step}$ is the latency per denoising step. Meanwhile, $z_i + x_i + y_i = 1$ means that only one case applies per request.

\subsection{Cache Maintenance Policy: \texttt{LCU}}
  \label{sec:lcu}
  The CacheGenius generates substantial images while providing services to users. Whenever a new image is generated, it is encoded into a vector through the \texttt{embedding-generator} and inserted into the VDB as a reference for subsequent generations. The VDB preloads numerous images from public datasets during initialization, already occupying a portion of the storage capacity. While it is theoretically possible to store all data in the VDB until reaching the storage limit, excessive caching beyond a certain threshold degrades the VDB performance, leading to slower query speeds. Consequently, considering both storage costs and query efficiency, the VDB requires periodic updates. 
  \begin{algorithm}[h]
    \caption{Cache Maintenance Policy: \texttt{LCU}}
    \label{alg:lcu}
    \KwIn{VDBs $\{\mathcal{D}_1,...,\mathcal{D}_{|\mathbf{N}|}\}$, cache capacity $C_{\text{max}}$}
    \KwOut{Updated VDBs $\{\mathcal{D}'_1,...,\mathcal{D}'_{|\mathbf{N}|}\}$}
    
    \For{each node $k \in [1,{|\mathbf{N}|}]$}{
        $L_k \gets [\ ]$\;
        \For{each $v \in \mathcal{D}_k$}{
            $d \gets \|v - \mu_k\|_2$\;
            $L_k.\text{append}((v, d, k))$\;
        }
    }
    
    $L\gets \bigcup_{k=1}^{|\mathbf{N}|} L_k$ \Comment{Merge all node lists}
    Sort $L$ by $d$ in \textbf{descending order}\;
    
    \While{$\sum_{k=1}^{|\mathbf{N}|} |\mathcal{D}_k| > C_{\text{max}}$}{
        $(v, d, k) \gets L.\text{pop}(0)$\;
        $\mathcal{D}_k \gets \mathcal{D}_k \setminus \{v\}$ \Comment{Remove from original VDB}
    }
    
    \Return $\{\mathcal{D}'_1,...,\mathcal{D}'_{|\mathbf{N}|}\}$
  \end{algorithm}

    Traditional policies like Least Recently Used (LRU) or Least Frequently Used (LFU), and First Input First Output (FIFO) exhibit significant limitations by ignoring semantic relationships, potentially discarding valuable early high-relevance vectors. These policies fail to proactively measure semantic correlations between vectors, hindering the maintenance of semantic clusters and efficient querying. To overcome these problems, the cache maintenance policy \texttt{LCU} is proposed. Compared to traditional LRU and LFU policies, \texttt{LCU} innovatively integrates cache maintenance with semantic distribution. It establishes a semantic relevance evaluation module between cached images and the current VDB. The module operates by calculating the center of the data distribution in the VDB and prioritizing the removal of samples that exhibit the greatest Euclidean distance deviation relative to their center, which fundamentally constructs an outlier detection framework based on the data distribution. These outliers often contain mixed semantic concepts that provide limited reference value for image generation (illustrated as outlier samples within red circles in Figure \ref{fig:tsne}). Algorithm \ref{alg:lcu} details the \texttt{LCU} workflow.


    \begin{figure}[t]
      \centering
      \includegraphics[width=\linewidth]{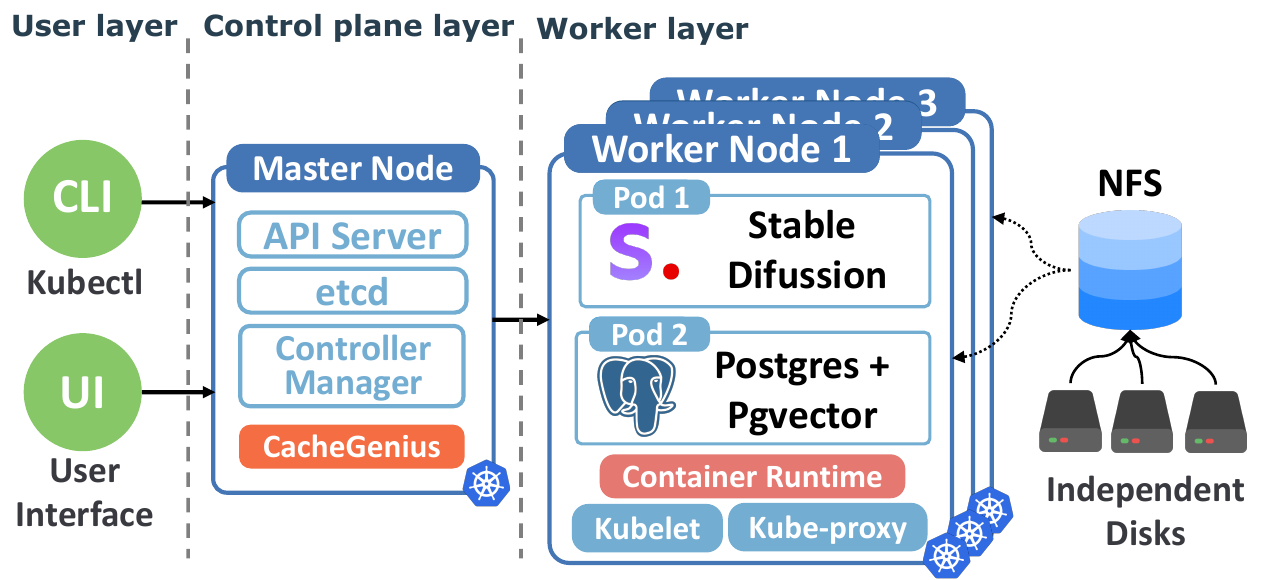}
      \caption{System architecture of CacheGenius}
      \label{fig:system}
    \end{figure}

    \section{Implementation}
    CacheGenius is deployed in a distributed, heterogeneous edge environment connected via a wired Ethernet network. The cluster comprises four PCs that serve as edge devices: one Master Node and three Worker Nodes. The Master Node is equipped with an NVIDIA RTX 4090 D GPU, and the Worker Nodes are equipped with NVIDIA RTX 4090 D, RTX 3090, and RTX 2070 SUPER GPUs, respectively. The system implements a VDB using pgvector, a PostgreSQL extension that provides efficient vector indexes and distance computations. The overall environment follows a layered architecture, orchestrated by Kubernetes, as shown in Figure \ref{fig:system}.
    
    \textbf{Control plane layer}: The control plane layer, centered on the Master Node, orchestrates global resource scheduling and cluster management. The API server serves as the control hub, authenticating and processing user requests.  The etcd database is a distributed key-value storage engine responsible for persistently storing cluster configurations, Pod definitions, and node status metadata. The Controller Manager maintains system-state consistency through continuous control loops. CacheGenius operates within the Master Node, sequentially invoking modules including the \texttt{embedding-generator} and \texttt{request-scheduler} upon user request reception, ultimately scheduling tasks to Worker Nodes for image generation. To handle large-scale concurrent user requests and maintain load balance across heterogeneous nodes, we implement an asynchronous task queue that decouples request intake from image generation.
    
    \textbf{Worker layer}: The worker layer aggregates multiple Worker Nodes into a physical compute resource pool. Each node uses the Kubelet to communicate with the Master Node. Kube-proxy maintains dynamic network rules at the node level, and the containerd runtime interfaces with Kubernetes via the Container Runtime Interface (CRI). We deploy two Pods per Worker Node: one Pod hosts a CUDA-accelerated Stable Diffusion service for image generation, while the other runs PostgreSQL with the pgvector extension to store vector embeddings of generated images and captions. A significant dependency management challenge is that the official Stable Diffusion image on Docker Hub \cite{dockerhub} omits critical CUDA-related libraries (e.g., xFormers \cite{xFormers2022}), adding 25-40 seconds of dependency downloads to each Pod initialization. We address this by rebuilding the Docker image with preinstalled dependencies, reducing cold-start latency to under 10 seconds. A 500GB NFS via PersistentVolume provides shared storage for model files, generated images, and VDB assets.

    \begin{figure}[t]
      \centering
      \includegraphics[width=0.5\textwidth]{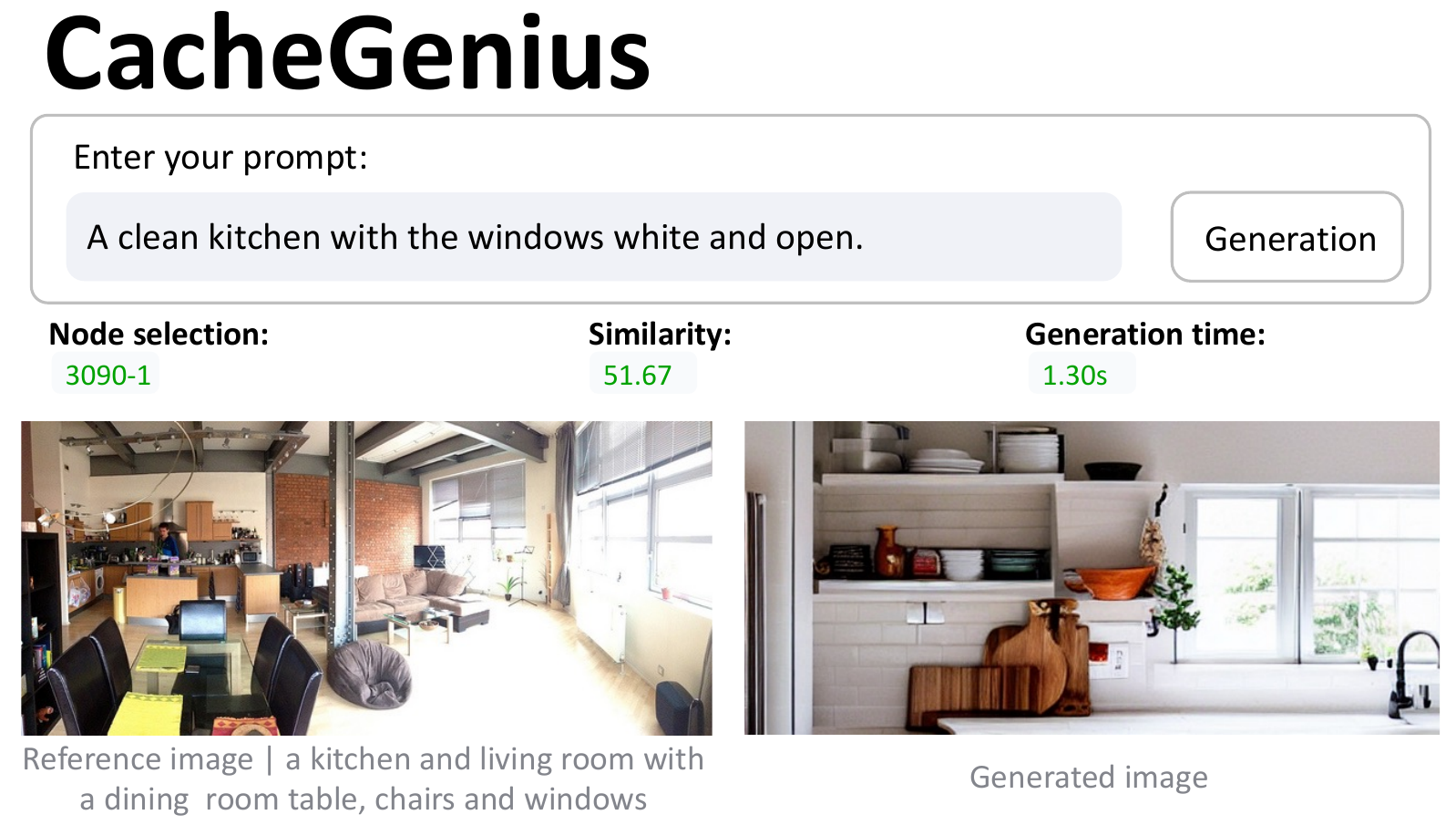}   
      \caption{Web UI of CacheGenius}
      \label{fig:ui}
    \end{figure}
    
    \textbf{User layer}: The user layer provides a Command-Line Interface (CLI) tool that supports batch submissions for automated testing and operational scripting. To empower non-technical users, we design an intuitive web User Interface (UI), as illustrated in Figure \ref{fig:ui}. The UI streams real-time generation results and VDB retrievals over a bidirectional WebSocket connection \cite{fette2011websocket}. The WebSocket implementation ensures low-latency communication between the frontend and backend. Both access channels (CLI and UI) share a semantic-aware caching infrastructure while maintaining distinct optimization strategies. Furthermore, Kubectl facilitates application deployment, Pod monitoring, and cluster management.
    

    \section{Evaluation}
    
    \textbf{Overview}: This section assesses CacheGenius by conducting extensive experiments within a distributed heterogeneous edge computing environment. The evaluation employs a comprehensive set of similarity and quality metrics, including CLIP \cite{radford2021learning}, PickScore \cite{kirstain2023pick}, Inception Score \cite{salimans2016improved}, and Fréchet Inception Distance (FID) \cite{heusel2017gans}, which together quantify the semantic alignment and visual performance of the generated images. CacheGenius is evaluated against several baselines, including GPT-CACHE \cite{bang2023gptcache}, PINECONE \cite{pinecone2023stable}, NIRVANA \cite{agarwal2024approximate}, as well as compressed models like SD-Tiny \cite{kim2023bksdm}. The results demonstrate that CacheGenius maintains image generation quality competitive with Stable Diffusion \cite{Rombach2022} while substantially improving computational efficiency. 
    
    In terms of system efficiency, CacheGenius shows a substantial reduction in generation latency and computational costs. It achieves a 41\% latency reduction compared to Stable Diffusion, while the \texttt{LCU} cache maintenance policy outperforms traditional caching methods by maintaining higher cache hit rates. The evaluation also details the impact of \texttt{request-scheduler} and \texttt{prompt-optimizer} components, revealing improved responsiveness and better image quality. Furthermore, cost analysis highlights the economic advantages of CacheGenius, achieving an approximate 48\% reduction in resource expenditure by minimizing redundant computations.

    \textbf{Baselines}: To compare the performance, several baselines are conducted. The details are as follows.
    \begin{itemize}[noitemsep, leftmargin=*, topsep=0pt]
      \item \textbf{GPT-CACHE} \cite{bang2023gptcache}: Retrieves image for the closest prompt based on BERT embedding similarity. Otherwise, it generates an image from random noise.
      \item \textbf{PINECONE} \cite{pinecone2023stable}: Retrieves an image for the closest prompt based on CLIP embedding similarity. Otherwise, it generates an image from random noise.
      \item \textbf{NIRVANA} \cite{agarwal2024approximate}: Uses approximate caching to reuse intermediate noise states from previous text-to-image generations.
      \item \textbf{CacheGenius w/o CMP}: CacheGenius without cache maintenance policy: \texttt{LCU}.
      \item \textbf{CacheGenius w/o RS}: CacheGenius without \texttt{request} \texttt{-scheduler} component.
      \item \textbf{SD-Tiny} \cite{kim2023bksdm}: SD-Tiny is an architecturally compressed Stable Diffusion model with 0.5 billion parameters for efficient general-purpose text-to-image generation.
      \item \textbf{Stable Diffusion} \cite{Rombach2022}: An open-source text-to-image DM that generates high-quality images from text prompts through iterative latent denoising.
    \end{itemize}

    \textbf{Evaluation metrics}: We evaluate CacheGenius on various metrics that cover both similarity and quality aspects.
    \begin{enumerate}[label=\textbf{\arabic*.},noitemsep, leftmargin=*, topsep=0pt]
      \item \textbf{Similarity metrics}:
      \begin{itemize}[noitemsep, leftmargin=*, topsep=0pt]
        \item \textbf{CLIP Score} \cite{radford2021learning}: Measures text-image alignment by computing cosine similarity between CLIP model embeddings of text prompts and generated images.
        \item \textbf{PickScore} \cite{kirstain2023pick}: Evaluates human preference alignment by scoring images based on a model trained on human-ranked data for aesthetic and semantic quality.
      \end{itemize}
      \item \textbf{Quality metrics}:
      \begin{itemize}[noitemsep, leftmargin=*, topsep=0pt]
        \item \textbf{Inception Score} \cite{salimans2016improved}: Quantifies image quality and diversity using the entropy of class predictions from a pre-trained Inception-v3 network \cite{szegedy2015}.
        \item \textbf{FID} \cite{heusel2017gans}: Assesses realism by calculating the Wasserstein-2 distance between feature distributions of real and generated images via Inception-v3 embeddings.
      \end{itemize}
    \end{enumerate}

    \textbf{Base diffusion models}: We employ Stable Diffusion v1.5 and v2.1 as the base models for image generation, with 1.04 billion and 1.29 billion parameters, respectively \cite{Rombach2022}. By default, Stable Diffusion generates $512\times512$ resolution images. The model parameters downloaded from Huggingface \cite{huggingface} are encapsulated into a Docker image to maintain experimental environment consistency.

    \textbf{Dataset}. Our reference image corpus is compiled from three captioned image sources: COCO \cite{lin2014microsoft}, DiffusionDB \cite{wangDiffusionDB}, and Flickr30k \cite{flickrentitiesijcv}. COCO provides 123K reference images. Each COCO image has five associated captions, and unless otherwise noted, we use the first caption. DiffusionDB offers text-to-image generation paired with user-supplied prompts. Flickr30k contains 31,000 photos, each with five human-written captions. For all experiments, image generation prompts are sourced from captions in the COCO 2017 validation images subset.

    \begin{figure*}[t]
      \centering
      \includegraphics[width=\linewidth]{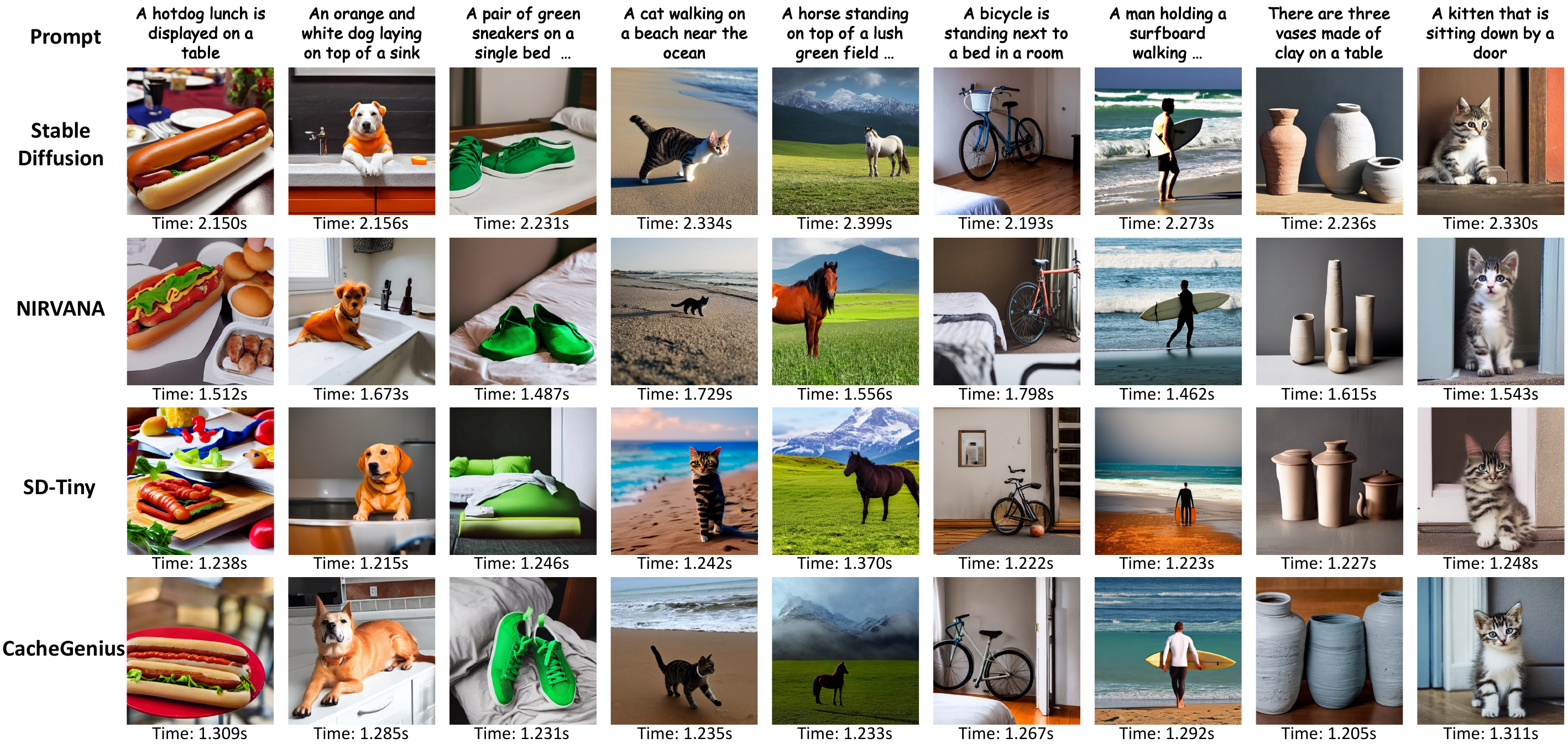}
      \caption{Real examples of generated images by Stable Diffusion, NIRVANA, SD-Tiny, and CacheGenius}
      \label{fig:compare}
    \end{figure*}

    \subsection{Overall Performance}
    
    \textbf{Quantitative generation performance}: Table \ref{tab:quality} highlights CacheGenius advancements across quantitative image evaluation metrics. Compared to retrieval-based baselines, GPT-CACHE and PINECONE, directly reuse images from the most similar cached prompt, leading to degraded similarity and quality metrics. This gap is critical in production settings, where user-facing text-to-image systems demand both precision and consistency. CacheGenius and its variants achieve substantial improvements in all metrics, underscoring their ability to address the limitations of retrieval-based baselines. NIRVANA reuses intermediate noise states via approximate caching, outperforms retrieval baselines, but falls short of CacheGenius. NIRVANA relies on long-running cached intermediate states, and its generation performance remains suboptimal during the initial stages when substantial intermediate states are still lacking.

    The impact of CacheGenius components is evident in its variants. CacheGenius w/o CMP and CacheGenius w/o RS reveal that removing the cache maintenance policy and \texttt{request-scheduler} modules slightly reduces performance. This highlights the complementary role of these modules in enhancing generation quality while also demonstrating that CacheGenius can achieve excellent performance in single-node scenarios or environments requiring no scheduling. It can be seen that CacheGenius even approaches the performance of Stable Diffusion, indicating that its outputs are nearly indistinguishable from the full model. Inception Score and FID comparisons further validate the robustness of CacheGenius. It outperforms all baselines except Stable Diffusion. Even SD-Tiny lags behind CacheGenius, emphasizing that model compression sacrifices quality, whereas CacheGenius retains efficacy through caching.

    \textbf{Generated images}: Figure \ref{fig:compare} demonstrates that CacheGenius achieves an optimal balance between efficiency and quality in image generation. As shown in the visual comparisons, CacheGenius produces images with comparable visual performance to those generated by Stable Diffusion, preserving intricate details and semantic alignment with textual prompts. Meanwhile, its generation latency closely matches that of SD-Tiny. Across diverse scenarios, including animals, plants, foods, and objects, both CacheGenius and Stable Diffusion produce visually equivalent outputs in terms of image and text correlation, detail preservation, and contextual coherence. This parity in visual performance confirms that CacheGenius successfully replicates Stable Diffusion generation capabilities while leveraging its hybrid image generation system to accelerate denoising, maintaining critical aspects.
    
    \begin{table}[t]
      \centering
      \resizebox{\columnwidth}{!}{%
      \begin{tabular}{ccccc}
        \toprule
        \multicolumn{1}{c|}{\multirow{2}{*}{\textbf{Method}}} & \multicolumn{2}{c|}{\textbf{Similarity Metrics}}                  & \multicolumn{2}{c}{\textbf{Quality Metrics}}  \\ \cline{2-5} 
        \multicolumn{1}{c|}{}                                 & CLIP Score $\uparrow$ & \multicolumn{1}{c|}{PickScore $\uparrow$} & Inception Score $\uparrow$ & FID $\downarrow$ \\ \midrule
        \multicolumn{5}{c}{\textbf{Stable Diffusion v1.5}}                                                                                 \\ \midrule
        \multicolumn{1}{c|}{Original}             & 31.44          & \multicolumn{1}{c|}{21.83}          & 31.59          & 17.03          \\
        \multicolumn{1}{c|}{GPT-CACHE}            & 24.60          & \multicolumn{1}{c|}{19.59}          & 27.85          & 21.86          \\
        \multicolumn{1}{c|}{PINECONE}             & 26.76          & \multicolumn{1}{c|}{20.06}          & 27.01          & 22.65          \\
        \multicolumn{1}{c|}{NIRVANA}              & 29.53          & \multicolumn{1}{c|}{20.75}          & 30.41          & 19.50          \\
        \multicolumn{1}{c|}{CacheGenius w/o CMP}  & 30.47          & \multicolumn{1}{c|}{21.20}          & 30.63          & 19.09          \\
        \multicolumn{1}{c|}{CacheGenius w/o RS}   & 30.46          & \multicolumn{1}{c|}{21.26}          & 28.89          & 19.91          \\
        \multicolumn{1}{c|}{SD-Tiny}              & 29.30          & \multicolumn{1}{c|}{20.69}          & 30.36          & 19.80          \\
        \multicolumn{1}{c|}{\textbf{CacheGenius}} & \textbf{30.51} & \multicolumn{1}{c|}{\textbf{21.28}} & \textbf{31.12} & \textbf{18.83} \\ \midrule
        \multicolumn{5}{c}{\textbf{Stable Diffusion v2.1}}                                                                                 \\ \midrule
        \multicolumn{1}{c|}{Original}             & 32.37          & \multicolumn{1}{c|}{19.73}          & 31.46          & 16.29          \\
        \multicolumn{1}{c|}{GPT-CACHE}            & 25.78          & \multicolumn{1}{c|}{17.61}          & 27.14          & 19.34          \\
        \multicolumn{1}{c|}{PINECONE}             & 25.58          & \multicolumn{1}{c|}{18.36}          & 27.33          & 19.33          \\
        \multicolumn{1}{c|}{NIRVANA}              & 30.17          & \multicolumn{1}{c|}{19.12}          & 30.39          & 18.47          \\
        \multicolumn{1}{c|}{CacheGenius w/o CMP}  & 31.59          & \multicolumn{1}{c|}{19.34}          & 31.09          & 17.87          \\
        \multicolumn{1}{c|}{CacheGenius w/o RS}   & 31.57          & \multicolumn{1}{c|}{19.42}          & 30.85          & 18.18          \\
        \multicolumn{1}{c|}{\textbf{CacheGenius}} & \textbf{31.60} & \multicolumn{1}{c|}{\textbf{19.57}} & \textbf{31.22} & \textbf{16.87} \\ \bottomrule
        \end{tabular}%
        }
      \caption{Comparative evaluation of image generation models using similarity and quality metrics}
      \label{tab:quality}
      \end{table}

    \begin{figure}[ht] 
      \centering
        \centering
        \begin{subfigure}[b]{0.235\textwidth} 
          \includegraphics[width=\linewidth]{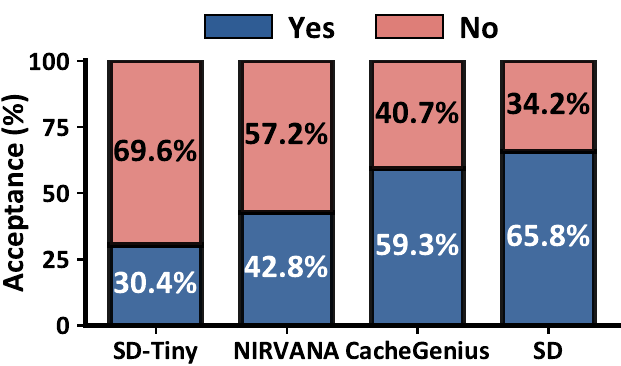}
          \caption{Acceptance of the quality of the generated image}
          \label{fig:Acceptance}
        \end{subfigure}
        \hfill
        \begin{subfigure}[b]{0.235\textwidth} 
          \includegraphics[width=\linewidth]{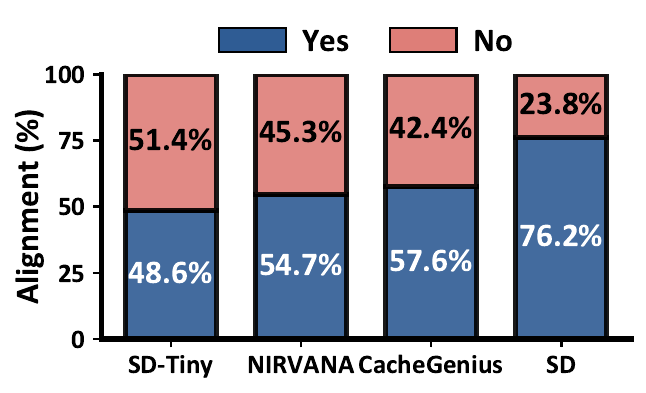}
          \caption{Alignment of the prompt and the generated image}
          \label{fig:Alignment}
        \end{subfigure}
        \caption{User survey results}
        \label{fig:survey}
    
    \end{figure}

    \begin{figure*}[t]
      \centering
      \begin{minipage}[t]{0.32\textwidth}
        \centering
        \includegraphics[width=\linewidth]{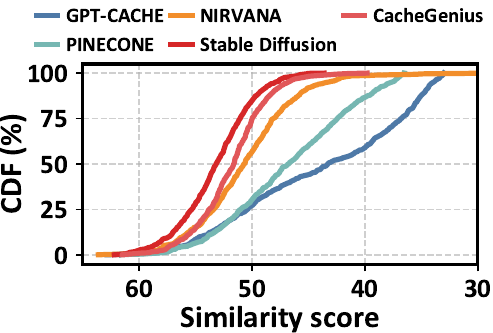}
        \caption{CDF of similarity score}
        \label{fig:cdf}
      \end{minipage}
      \hfill
      \begin{minipage}[t]{0.32\textwidth}
        \centering
        \includegraphics[width=\linewidth]{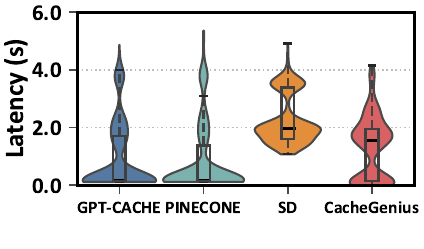}
        \caption{Distribution of the response latencies for approaches}
        \label{fig:vio}
      \end{minipage}
      \hfill
      \begin{minipage}[t]{0.31\textwidth}
        \centering
        \includegraphics[width=\linewidth]{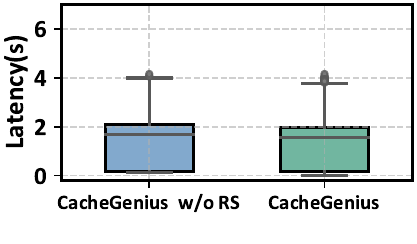}
          \caption{Comparison of generation latency on \texttt{request-scheduler}}
      \label{fig:so_time}
      \end{minipage}
    \end{figure*}

    \textbf{Human evaluation}: We conduct a survey with 40 participants to evaluate the quality and semantic alignment of the images generated by SD-Tiny, NIRVANA, Stable Diffusion, and CacheGenius. Each participant is randomly assigned 3 images from each model (12 images total) and asked to assess whether the image quality is acceptable and whether the generated image semantically aligns with the prompt. The results shown in Figure \ref{fig:survey} reveal that CacheGenius achieved a 59.3\% acceptance rate for quality and a 57.6\% alignment rate between prompts and images. SD-Tiny exhibited the lowest alignment rate, highlighting the trade-off between model compression and output quality. This experiment demonstrates the effectiveness of CacheGenius in optimizing latency-sensitive applications while maintaining user-perceived quality.

    In summary, CacheGenius performs well in terms of similarity and quality. It closely matches Stable Diffusion performance while surpassing retrieval and compressed models. The minor performance gaps between CacheGenius and its variants further underscore the necessity of its integrated components to achieve optimal results.
    
    \subsection{System Efficiency}
    \label{sec:efficiency}
    \textbf{CDF analysis of similarity score}: The Cumulative Distribution Function (CDF) plot in Figure \ref{fig:cdf} illustrates the cumulative distribution of similarity scores across different baselines, further validating the superiority of CacheGenius in preserving prompt-image alignment. Retrieval-based baselines exhibit flatter curves, with only about 20\% of requests achieving a similarity score above 50, indicating that they often fail to retrieve images matching the prompts, as their embedding-based matching struggles to adapt to diverse input variations. In contrast, CacheGenius demonstrates a steeper ascent, covering about 80\% of requests above a similarity score of 50. The proportion of CacheGenius with mid-range scores (50-55) closely matches that of Stable Diffusion. These patterns align with quantitative CLIP Score and PickScore, confirming that CacheGenius achieves optimal prompt-image alignment while maintaining computational efficiency.
    
    \textbf{Latency}: Table \ref{tab:time} compares the average latency and percentile-to-median ratios for four image generation methods. CacheGenius achieves an average latency of 1.32 seconds, outperforming Stable Diffusion but slightly lagging behind retrieval-based baselines GPT-CACHE and PINECONE. CacheGenius demonstrates superior consistency with significantly lower percentile-to-median ratios than baselines. In contrast, GPT-CACHE and PINECONE show high variability, indicating unstable performance.
    \begin{table}[h]
      \centering
      \resizebox{\columnwidth}{!}{
      \begin{tabular}{c|cccc}
      \toprule
      & \textbf{Latency(s)} & \textbf{$90^{th}$/median} & \textbf{$95^{th}$/median} & \textbf{$99^{th}$/median} \\ \midrule
      GPT-CACHE        & 0.91    & 13.71       & 21.74       & 22.92       \\
      PINECONE         & 0.78    & 14.31       & 21.76       & 23.87       \\
      NIRVANA         & 1.61    & 1.20       & 1.21       & 1.21       \\
      SD-Tiny        & 1.24    & 1.01       & 1.01       & 1.02       \\
      Stable Diffusion & 2.24    & 1.07        & 1.10        & 1.14        \\
      CacheGenius      & 1.32    & 1.09        & 1.12       & 1.18        \\ \bottomrule
      \end{tabular}}
      \caption{Average latency and $n^{th}$ percentile over median values of the response latencies for approaches}
      \label{tab:time}
      \end{table}

    Figure \ref{fig:vio} illustrates the latency distribution of image generation systems, including CacheGenius, Stable Diffusion, GPT-CACHE, and PINECONE. CacheGenius achieves a balanced trade-off between latency reduction and stability, leveraging reference image generation. While retrieval-based baselines exhibit lower median latencies, their long-tail latency peak highlights reliability limitations. 

    \textbf{Performance with request scheduler}: Figure \ref{fig:so_time} illustrates the comparative analysis of image generation latency between the CacheGenius system with and without the \texttt{request-scheduler} component. This figure emphasizes the critical role of the \texttt{request-scheduler} module in optimizing generation latency, combining quality-aware priority scheduling and historical query caching to ensure that tasks are scheduled to the most appropriate node. This improvement highlights the effectiveness of CacheGenius in employing scheduling strategies to enhance real-time responsiveness in image generation workflows.

    \begin{figure}[t]
      \centering
      \begin{minipage}[t]{0.235\textwidth}
        \centering
        \includegraphics[width=\linewidth]{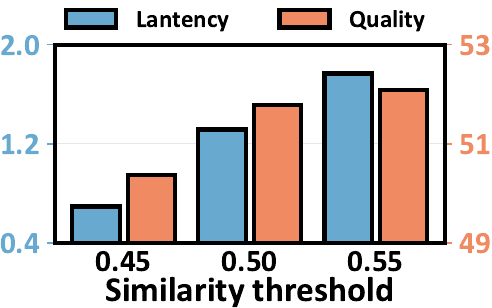}
        \captionof{figure}{Comparison of generation latency and generation quality of similarity threshold}
        \label{fig:para}
      \end{minipage}
      \hfill
      \begin{minipage}[t]{0.235\textwidth}
        \centering
        \includegraphics[width=\linewidth]{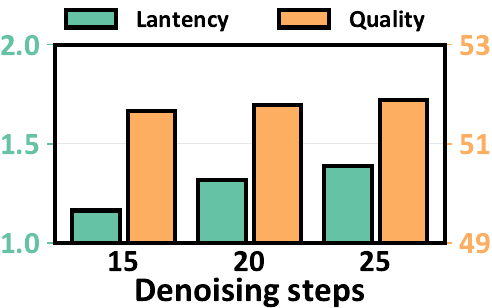}
        \captionof{figure}{Comparison of generation latency and generation quality of denoising steps}
        \label{fig:step}
      \end{minipage}
      \label{fig:exp2}
    \end{figure}

    \begin{figure*}[t]
      \centering
      \begin{minipage}[t]{0.32\textwidth}
        \centering
        \includegraphics[width=\linewidth]{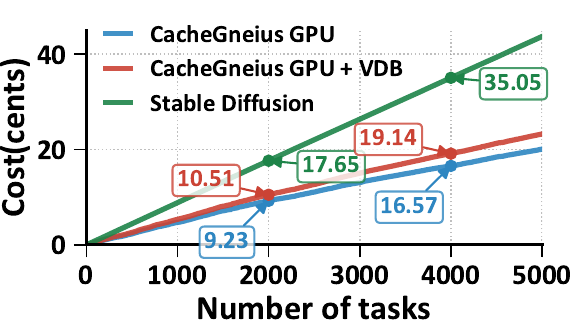}
        \captionof{figure}{Cost comparison of model components against a stream of queries over time}
        \label{fig:cost}
      \end{minipage}
      \hfill
      \begin{minipage}[t]{0.32\textwidth}
        \centering
        \includegraphics[width=\linewidth]{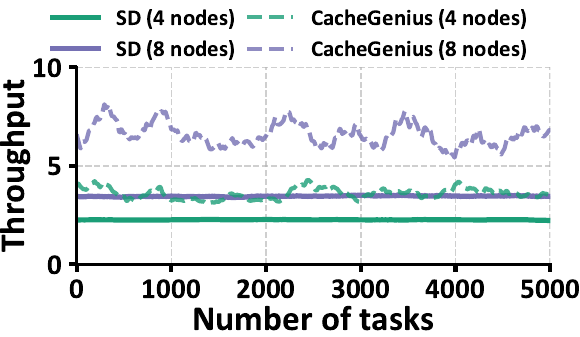}
        \captionof{figure}{Throughput comparison of models against a stream of queries over time}
        \label{fig:throughput}
      \end{minipage}
      \hfill
      \begin{minipage}[t]{0.32\textwidth}
        \centering
        \includegraphics[width=\linewidth]{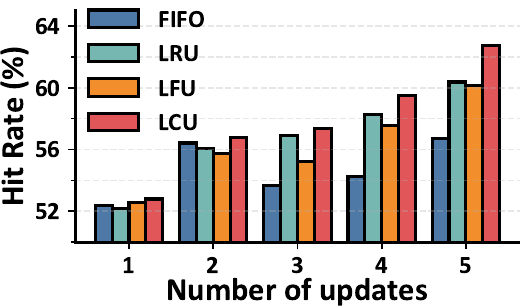}
        \captionof{figure}{Hit rate after five updates using different policies}
        \label{fig:policy}
      \end{minipage}
      \label{fig:exp3}
    \end{figure*}

    \textbf{Performance with different similarity threshold}: Figure \ref{fig:para} demonstrates the trade-off between generation efficiency and output quality under different similarity thresholds. 
    The multi-threshold chart shows that lower thresholds reduce latency through efficient VDB retrieval for image generation, but lower quality due to poor semantic alignment. Higher thresholds increase latency from failed retrievals requiring direct text-to-image generation, yet achieve the highest quality via full alignment. 
    The data suggest an optimal threshold of around 0.5 for practical applications, balancing acceptable generation processes with quality preservation. 
    
    \textbf{Performance with different denoising steps}: Figure \ref{fig:step} illustrates the trade-off between generation quality and latency under different denoising steps. When leveraging reference images retrieved from the VDB for image-to-image generation, increasing denoising steps improves output quality by refining structural and semantic alignment with the prompt. However, this enhancement incurs linearly increasing latency due to the diffusion processes. Based on validation, CacheGenius adopts 20 denoising steps as the default configuration to achieve a balance between quality and latency. 

    \textbf{Performance with prompt optimizer}: Table \ref{tab:sort} quantifies the impact of the \texttt{prompt-optimizer} component on image generation quality, measured by Inception Score and FID. Since the prompt is different after being processed by \texttt{prompt-optimizer}, it is impossible to compare the similarity scores of text and images. Only the quality of the generated image itself is compared. The improvement in Inception Score and reduction in FID demonstrate that a structured prompt enhances semantic precision in generated images.
    
    \begin{table}[t]
      \centering
      \resizebox{\columnwidth}{!}{
      \begin{tabular}{c|ccc}
      \toprule
                 & \textbf{Inception Score} $\uparrow$ & \textbf{FID} $\downarrow$ & \textbf{Latency(s)}   \\ \midrule
      CacheGenius w/o PO & 30.50 & 19.03 & 1.20\\
      CacheGenius        & 31.12  & 18.83 & 1.32\\ \bottomrule
      \end{tabular}}
      \caption{Ablation study on \texttt{prompt-optimizer}}
      \label{tab:sort}
      \end{table}
    

    \textbf{Cost analysis of image generation}: To evaluate the economic benefits of CacheGenius compared to Stable Diffusion, we conduct a detailed cost analysis based on AutoDL GPU pricing \cite{autodl}. CacheGenius incurs two primary costs, GPU usage for the diffusion process and VDB storage and query operations. For GPU costs, we deploy CacheGenius across multiple instances, NVIDIA RTX 4090 D GPU (\$0.28/hour), NVIDIA RTX 3090 GPU (\$0.23/hour), and NVIDIA RTX 2070 GPU (\$0.084/hour), depending on task requirements. The VDB cost is \$0.12 per hour for storage and query operations. Stable Diffusion relies solely on GPU resources.
    
    In Figure \ref{fig:cost}, we aggregate costs across 5000 user tasks to compare total expenses. Stable Diffusion cost reflects uninterrupted GPU consumption without caching benefits. The results demonstrate that CacheGenius achieves a 48\% reduction in overall cost compared to Stable Diffusion. This dramatic efficiency stems from CacheGenius ability to minimize redundant computations through caching, offsetting the marginal VDB expenses. Such cost-effectiveness underscores CacheGenius practicality for large-scale, real-world applications, where computing resource saving is critical.

    \textbf{Throughput across different numbers of edge nodes}: The comparative analysis of the throughput achieved by CacheGenius and Stable Diffusion across varying edge node configurations is presented in Figure \ref{fig:throughput}. To assess system performance, we expand the edge environment to include four additional nodes, integrating one NVIDIA RTX 4090 D GPU and three NVIDIA RTX 2070 SUPER GPUs. Throughput is defined as the number of images generated per second by the overall system. The results indicate that both models exhibit improvements in throughput with the addition of nodes, but CacheGenius consistently outperforms Stable Diffusion. When running with four nodes, CacheGenius shows that the throughput is close to that of the eight nodes of Stable Diffusion. This performance improvement reflects CacheGenius significant reduction in image generation latency, validating its image generation efficiency under high-load scenarios.

    \textbf{Performance with \texttt{LCU}}: We evaluate the effectiveness of the \texttt{LCU} against conventional caching policies, including LRU, LFU, and FIFO. This experiment is conducted under identical workload generator configurations with cache initialization. As shown in Figure \ref{fig:policy}, \texttt{LCU} achieves superior hit rates compared to LRU, LFU, and FIFO after 5 cache updates. The performance superiority arises from the semantic-driven eviction logic embedded in \texttt{LCU}, which retains vectors consistent with clustered VDB semantics and discards distributional outliers. 
    FIFO with linear time-based removal, LRU with recency prioritization, and LFU with frequency prioritization demonstrate the lowest efficiency due to inherent neglect of semantic relevance. 
    These results confirm that eviction via \texttt{LCU} is indispensable to maximize caching efficiency in the image generation workflow.

    \textbf{Impact of reference image similarity}: Our method leverages semantic matching between the prompt and reference images. To validate its necessity, we introduce two baselines: Random (a reference image uniformly sampled from the dataset) and Wrong (a hard negative whose semantics are intentionally mismatched with the prompt). As shown in Table \ref{tab:ref}, using correct reference images consistently outperforms random or wrong references across both CLIP Score and PickScore. While using a random reference can occasionally produce reasonable results for simple prompts, the average quality is significantly lower than with a semantically matched reference. This confirms that semantic similarity critically impacts generation quality.

    \begin{table}[t]
      \centering
      \small
      \begin{tabular}{c|cc}
      \toprule
      \textbf{Reference Image}  & \textbf{CLIP Score} $\uparrow$ & \textbf{PickScore} $\uparrow$ \\ \midrule
      Wrong    & 29.45                 & 18.86                \\
      Random   & 29.74                 & 19.01                \\
      Correct  & 31.60                 & 19.57                \\ \bottomrule
      \end{tabular}
      \caption{Quality of generation for different reference images}
      \label{tab:ref}
      \end{table}
    
    \textbf{Analysis of different embeddings}: Table \ref{tab:embeddings} summarizes the performance of different embeddings in terms of their CLIP Score and PickScore. The results indicate that using only BERT for text embedding obtained the lowest score. 
    However, when BERT is combined with the CLIP model for image embedding, there is a slight improvement. 
    The most notable performance improvement is observed when both text and image embeddings utilize the CLIP model, achieving the highest score. 
    This demonstrates that aligning both text and image embeddings with the CLIP model significantly enhances the overall image generation quality, indicating the effectiveness of CLIP embeddings in achieving better semantic matching.

    \begin{table}[t]
      \centering
      \resizebox{\columnwidth}{!}{
      \begin{tabular}{cc|cc}
        \toprule
        \textbf{Text Embedding} & \textbf{Image Embedding}  & \textbf{CLIP Score} $\uparrow$ & \textbf{PickScore} $\uparrow$ \\ \midrule
      BERT           & \textbackslash{} & 28.39      & 20.55     \\
      BERT           & CLIP             & 28.90      & 20.77     \\
      CLIP           & CLIP             & 30.51      & 21.28     \\ \bottomrule
      \end{tabular}
      }
      \caption{Quality of generation for different embeddings}
      \label{tab:embeddings}
      \end{table}

      \begin{figure}[t]
        \centering
        \includegraphics[width=\linewidth]{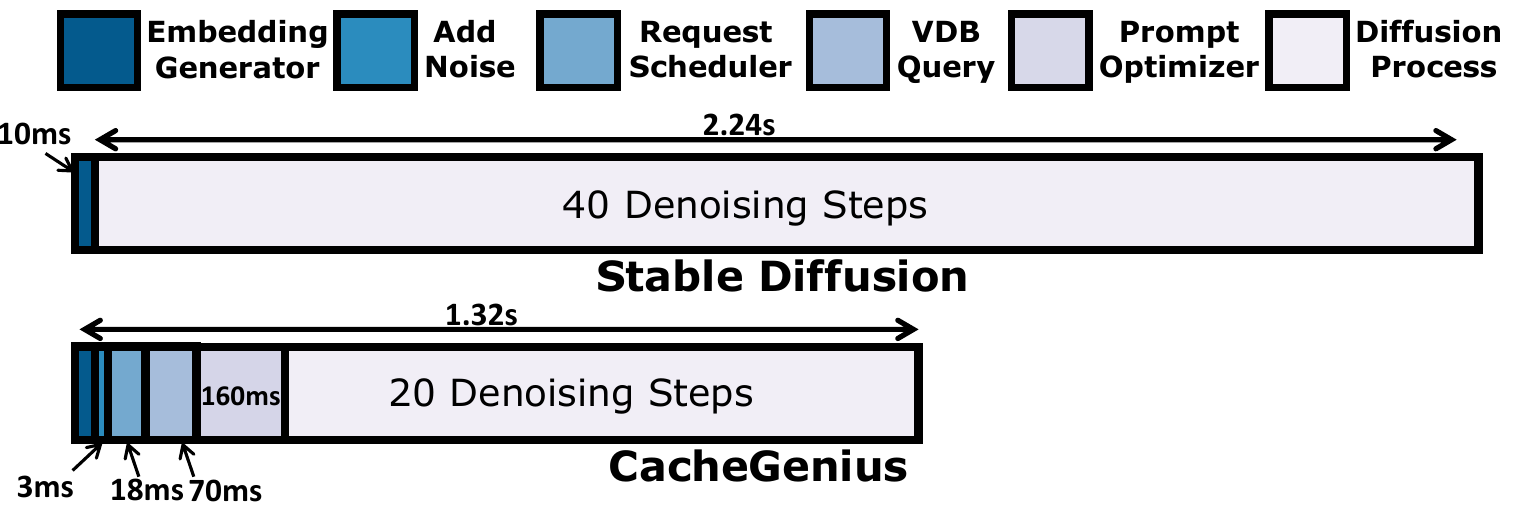}
        \caption{Latency taken by different components of Stable Diffusion and CacheGenius}
        \label{fig:time}
      \end{figure}

      \textbf{Latency distribution of CacheGenius}:
      Figure \ref{fig:time} presents the workflow and latency distribution of CacheGenius compared with Stable Diffusion. Both pipelines begin with the \texttt{embedding-generator}. CacheGenius then optimizes the prompt, schedules the prompt to the most relevant nodes, retrieves reference images via a VDB query, and completes generation by denoising the perturbed references. By contrast, Stable Diffusion omits these intermediate steps but must perform many more denoising iterations because it starts from pure random noise. The fundamental difference is that CacheGenius leverages cached references, whereas Stable Diffusion relies on iterative diffusion from random noise. Consequently, CacheGenius achieves a balanced latency-quality trade-off in image generation.

\begin{figure}[h]
        \centering
        \includegraphics[width=\linewidth]{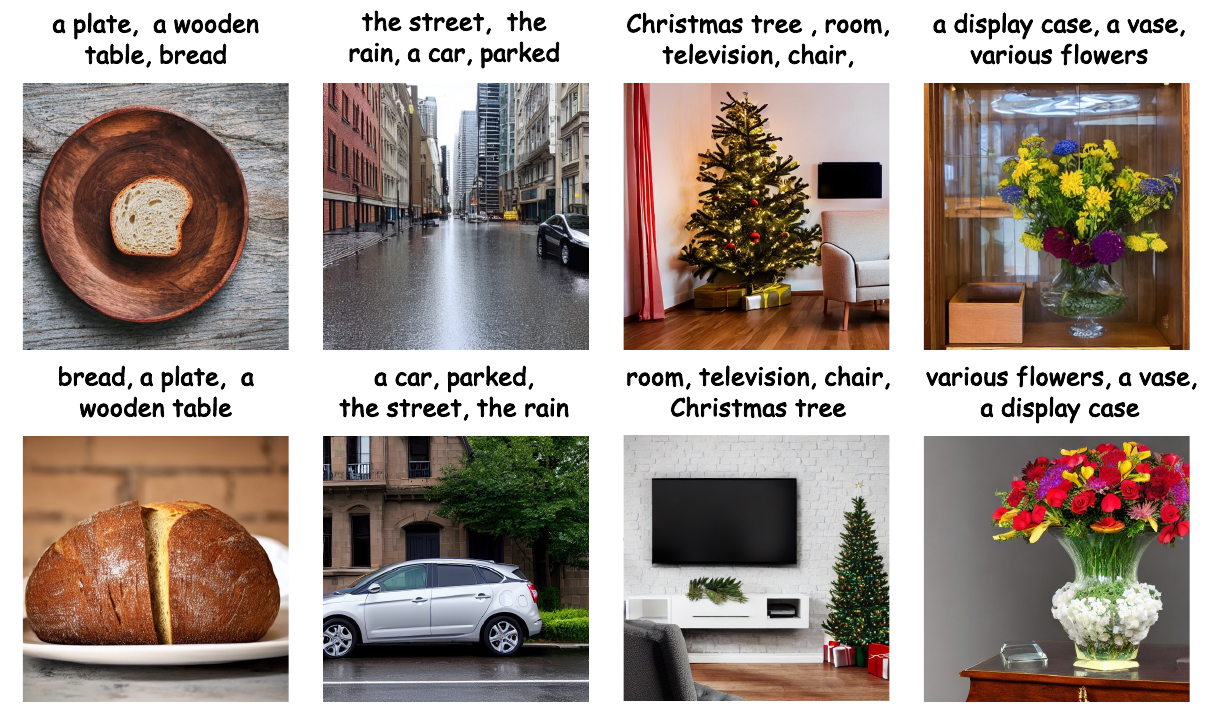}
        \caption{Generated images with different order prompts}
        \label{fig:split}
      \end{figure}

      \textbf{Analysis of Phrase Order Impact in Image Generation}:
      Figure \ref{fig:split} demonstrates how the sequential arrangement of phrases in image generation significantly influences visual composition. The figure presents four groups of prompts containing identical semantic phrases in varying orders, with each group generating two distinct images. It is observed that phrases positioned earlier in the prompt exert greater influence on the dominant element of generated images. For example, the prompt ``the street, the rain, a car, parked” emphasizes environmental context as a dominant element, while reversing the order ``a car, parked, the street, the rain” shifts focus to the car, reducing background detail prominence. Similarly, prioritizing ``Christmas tree” in the third group results in a centrally positioned tree, whereas placing ``room” first generates a broader scene with balanced elements. These findings align with the \texttt{prompt-optimizer} module in CacheGenius, which restructures prompts to align with diffusion model priors, thereby refining semantic alignment.
      
\section{Discussions}
\label{app:discuss}
\textbf{Complementarity with existing acceleration techniques}: Current approaches for reducing the generation latency of DMs primarily focus on model compression techniques, including distillation, pruning, and quantization. These methods often compromise output quality by reducing model complexity. The CacheGenius complements these approaches effectively, providing a mechanism to leverage semantic similarity and cached representations to accelerate generation without modifying the model architecture. Consequently, our system offers broader applicability by acting as a plug-in enhancement that can be integrated alongside compression methods, thereby enabling combined gains in speed and quality. 

\textbf{Extensibility to other iterative methods}: CacheGenius utilizes semantic similarity and cached representations to minimize redundant computations, demonstrating significant adaptability and potential for adaptation to other iterative generation frameworks beyond DM. For instance, the iterative refinement of architecture could be readily adapted to other sequential generative paradigms, such as sequential neural networks or alternative iterative sampling architectures. This flexibility suggests promising applications in domains such as video generation, where temporal coherence and large data continuity pose additional challenges. Caching partial computation results or reference states could significantly reduce redundancy in these contexts, and exploring such extensions forms a key direction for future research.

\textbf{Significance of the quality-speed trade-off}: 
While CacheGenius introduces a marginal degradation in output quality, this trade-off is highly palatable and beneficial in latency-critical applications. For example, in interactive use cases (e.g., metaverse scene generation, real-time design prototyping), reduced latency drastically improves user experience. Our human evaluation confirms that most of the users accepted CacheGenius outputs as qualitatively competitive despite minor differences. In resource-constrained edge environments, edge nodes have limited computing power. CacheGenius reduces GPU utilization by 48\% while preserving functional utility. For bulk image generation (e.g., dataset augmentation, content marketing), CacheGenius achieves near-linear throughput scaling, outperforming Stable Diffusion by 2.1$\times$ on 4-node clusters. The slight quality drop is negligible when balancing cost and scale.

    \section{Conclusions}
    We introduced CacheGenius, a hybrid image generation system that accelerates image generation. CacheGenius addresses the trade-off between output quality and computational efficiency on resource-constrained edge environments by combining semantic-aware caching with dynamic fusion of text-to-image and image-to-image workflows. Integrating a semantic-aware scheduling algorithm and the LCU policy ensures rapid retrieval of reference images aligned with evolving requests from mobile users. Evaluations in distributed edge environments demonstrate a 41\% reduction in latency for DM inference while maintaining competitive output quality. Extending CacheGenius to video generation could reduce redundant computations and improve efficiency through caching mechanisms, presenting a promising direction for future research.

\bibliographystyle{IEEEtran}
\bibliography{sample}










\newpage

\begin{IEEEbiography}[{\includegraphics[width=1in,height=1.25in,clip,keepaspectratio]{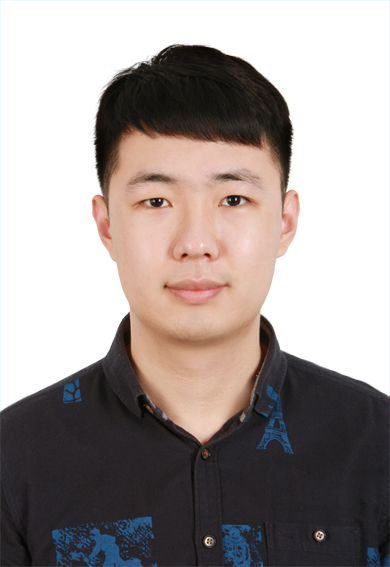}}]{Hanshuai Cui} received the B.S. degree from School of Information Science and Engineering, Qufu Normal University, China, in 2020. He is currently pursuing the Ph.D. degree in School of Artificial Intelligence, Beijing Normal University, China. His current research interests include mobile edge computing, resource allocation, and reinforcement learning.
\end{IEEEbiography}

\begin{IEEEbiography}[{\includegraphics[width=1in,height=1.25in,clip,keepaspectratio]{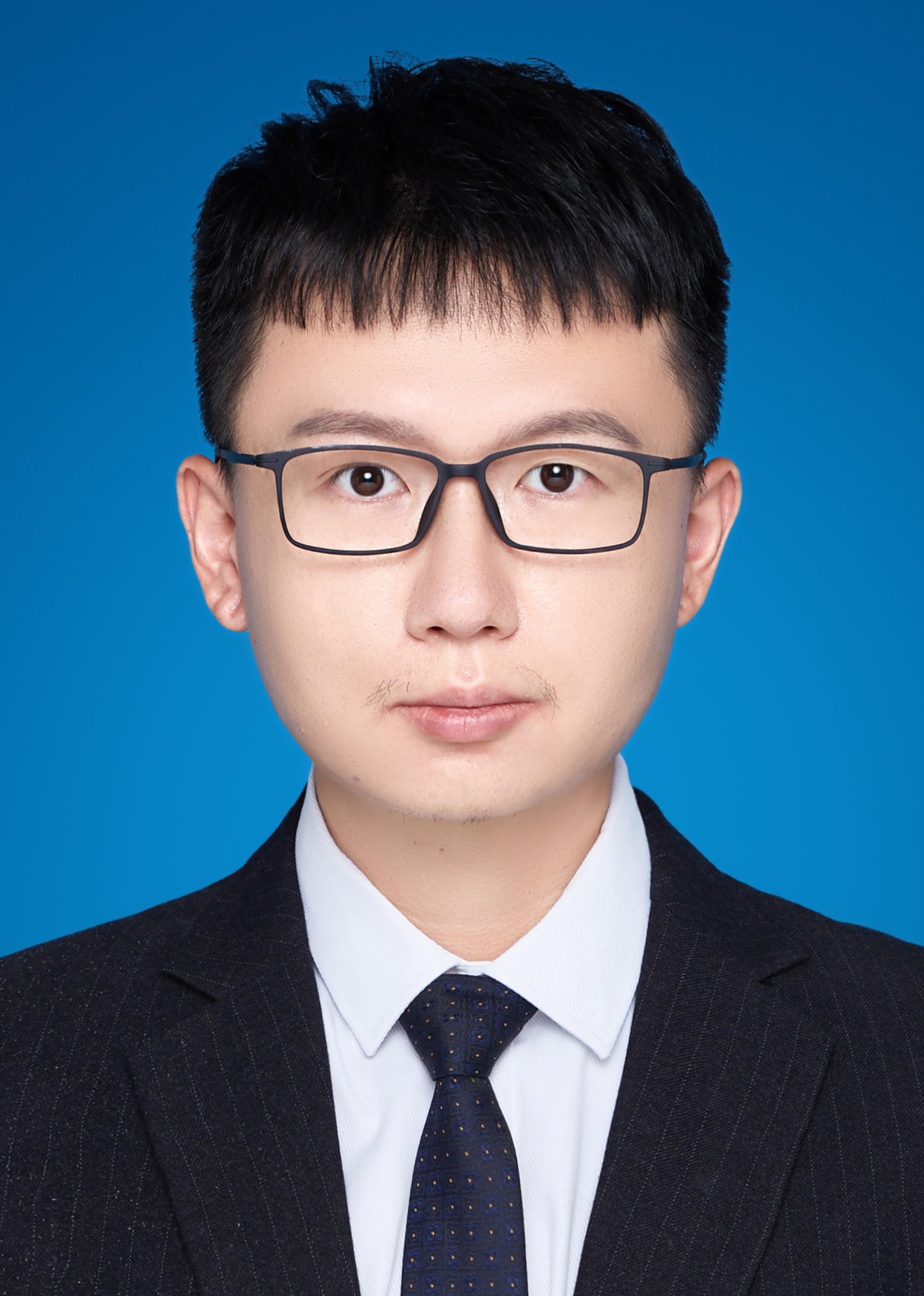}}]{Zhiqing Tang} received the B.S. degree from School of Communication and Information Engineering, University of Electronic Science and Technology of China, China, in 2015 and the Ph.D. degree from Department of Computer Science and Engineering, Shanghai Jiao Tong University, China, in 2022. He is currently an Associate Professor with the Advanced Institute of Natural Sciences, Beijing Normal University, China. His current research interests include edge computing, resource scheduling, and reinforcement learning.
\end{IEEEbiography}

\begin{IEEEbiography}[{\includegraphics[width=1in,height=1.25in,clip,keepaspectratio]{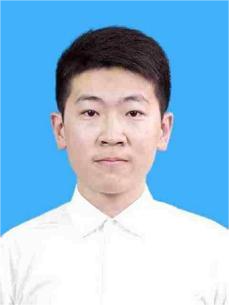}}]{Zhi Yao} received the B.S. degree from College of Electronic and Information Engineering, Shandong University of Science and Technology, China, in 2020, and the M.S. degree from South China Academy of Advanced Optoelectronics, South China Normal University, China, in 2023. He is currently pursuing the Ph.D. degree in School of Artificial Intelligence, Beijing Normal University, China. His current research interests include mobile edge computing, vector database, LLM request scheduling, and reinforcement learning.
\end{IEEEbiography}

\begin{IEEEbiography}[{\includegraphics[width=1in,height=1.25in,clip,keepaspectratio]{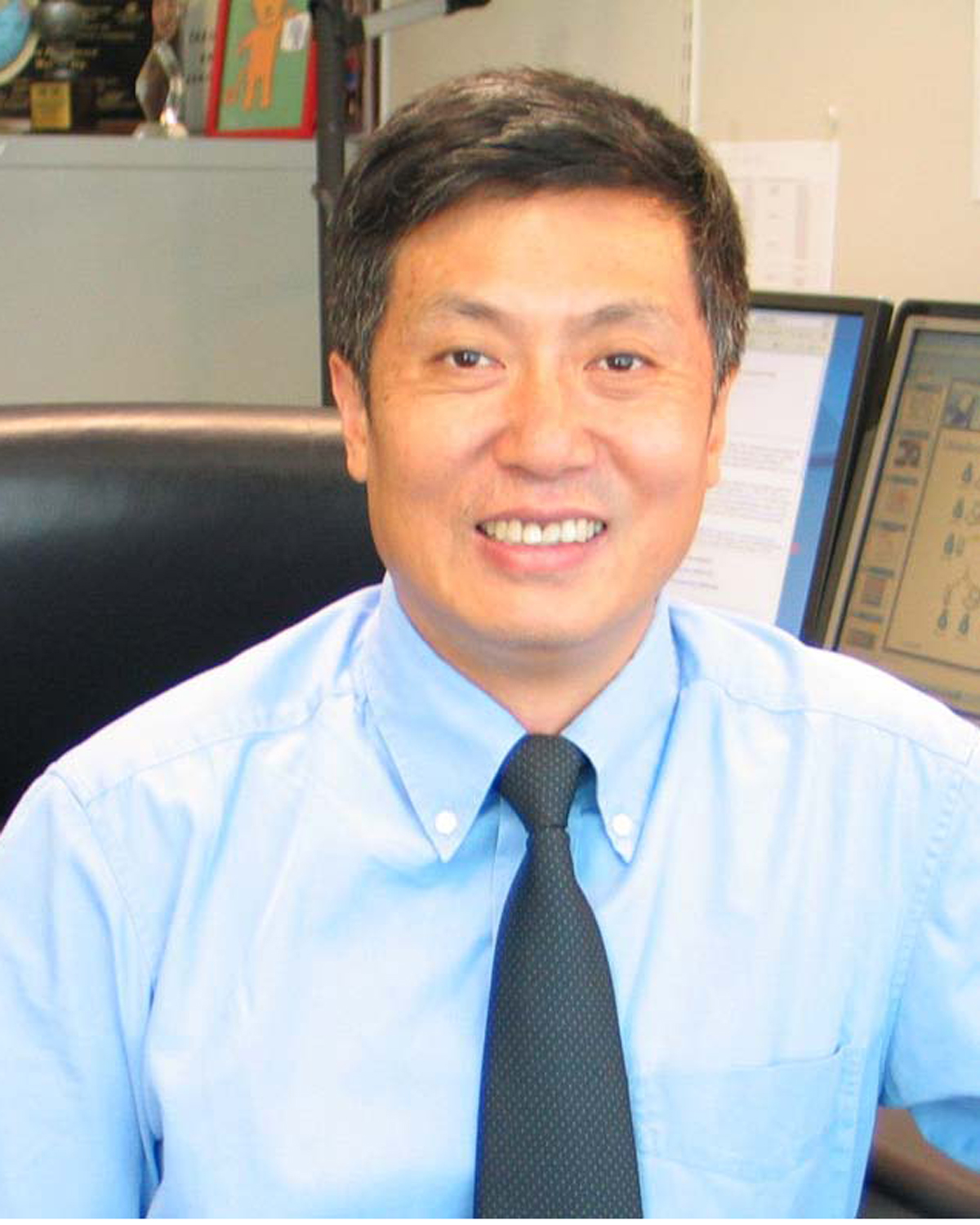}}]{Weijia Jia}
  (Fellow, IEEE) is currently the Director of Institute of Artificial Intelligence and Future Networking, and the Director of Super Intelligent Computer Center, Beijing Normal University at Zhuhai; also a Chair Professor at UIC, Zhuhai, Guangdong, China. He has served as the VP for Research at UIC in 6/2020-7/2024. Prior joining BNU, he served as the Deputy Director of State Key Laboratory of Internet of Things for Smart City at the University of Macau and Zhiyuan Chair Professor at Shanghai Jiaotong University, PR China. From 95-13, he worked in City University of Hong Kong as a professor. His contributions have been recoganized for the research of edge AI, optimal network routing and deployment; vertex cover; anycast and multicast protocols; sensors networking; knowledge relation extractions; NLP and intelligent edge computing. He has over 700 publications in the prestige international journals/conferences and research books and book chapters. He has received the best product awards from the International Science \& Tech. Expo (Shenzhen) in 2011/2012 and the 1st Prize of Scientific Research Awards from the Ministry of Education of China in 2017 (list 2), and  top 2\% World Scientists in Stanford-list (2020-2024) and many provincial science and tech awards. He has served as area editor for various prestige international journals, chair and PC member/keynote speaker for many top international conferences. He is the Fellow of IEEE and the Distinguished Member of CCF.
 \end{IEEEbiography}

\begin{IEEEbiography}[{\includegraphics[width=1in,height=1.25in,clip,keepaspectratio]{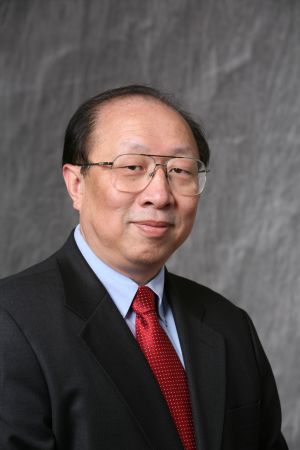}}]{Wei Zhao}(Fellow, IEEE) completed his undergraduate studies in physics at Shaanxi Normal University, China, in 1977, and received his MSc and PhD degrees in Computer and Information Sciences at the University of Massachusetts at Amherst in 1983 and 1986, respectively. Prof. Zhao has served important leadership roles in academic including the Chief Research Officer at the American University of Sharjah, the Chair of Academic Council at CAS Shenzhen Institute of Advanced Technology, the eighth Rector of the University of Macau, the Dean of Science at Rensselaer Polytechnic Institute, the Director for the Division of Computer and Network Systems in the U.S. National Science Foundation, and the Senior Associate Vice President for Research at Texas A\&M University. Prof. Zhao has made significant contributions to cyber-physical systems, distributed computing, real-time systems, and computer networks. His research results have been adopted in the standard of Survivable Adaptable Fiber Optic Embedded Network. Professor Zhao was awarded the Lifelong Achievement Award by the Chinese Association of Science and Technology in 2005. \end{IEEEbiography}

\vfill

\end{document}